\pdfoutput=1

\documentclass[12pt,a4paper]{article}

\usepackage{ifthen} 
\usepackage{multirow} 
\newboolean{pdflatex}
\setboolean{pdflatex}{true} 

\newboolean{articletitles}
\setboolean{articletitles}{true} 

\newboolean{uprightparticles}
\setboolean{uprightparticles}{false} 

\newboolean{inbibliography}
\setboolean{inbibliography}{false} 

\usepackage{microtype}
\usepackage{lineno}  
\usepackage{xspace} 
\usepackage{caption} 

\usepackage{graphicx}  
\usepackage{color}
\usepackage{colortbl}
\graphicspath{{./figs/}} 

\usepackage{amsmath} 
\usepackage{amssymb}
\usepackage{amsfonts}
\usepackage{upgreek} 

\newcommand*\patchAmsMathEnvironmentForLineno[1]{%
\expandafter\let\csname old#1\expandafter\endcsname\csname #1\endcsname
\expandafter\let\csname oldend#1\expandafter\endcsname\csname
end#1\endcsname
 \renewenvironment{#1}%
   {\linenomath\csname old#1\endcsname}%
   {\csname oldend#1\endcsname\endlinenomath}%
}
\newcommand*\patchBothAmsMathEnvironmentsForLineno[1]{%
  \patchAmsMathEnvironmentForLineno{#1}%
  \patchAmsMathEnvironmentForLineno{#1*}%
}
\AtBeginDocument{%
\patchBothAmsMathEnvironmentsForLineno{equation}%
\patchBothAmsMathEnvironmentsForLineno{align}%
\patchBothAmsMathEnvironmentsForLineno{flalign}%
\patchBothAmsMathEnvironmentsForLineno{alignat}%
\patchBothAmsMathEnvironmentsForLineno{gather}%
\patchBothAmsMathEnvironmentsForLineno{multline}%
\patchBothAmsMathEnvironmentsForLineno{eqnarray}%
}

\usepackage{hyperref}    
\usepackage[all]{hypcap} 

\def\lhcb {\mbox{LHCb}\xspace}

\def\MagUp {\mbox{\em Mag\kern -0.05em Up}\xspace}

\ifthenelse{\boolean{uprightparticles}}%
{

 \def\Pmu         {\ensuremath{\upmu}\xspace}                 
 \def\Pnu         {\ensuremath{\upnu}\xspace}                 
                  
 \def\Ppi         {\ensuremath{\uppi}\xspace}

 \def\Ppsi        {\ensuremath{\uppsi}\xspace}

 \def\PDelta      {\ensuremath{\Delta}\xspace}                 
 \def\PXi      {\ensuremath{\Xi}\xspace}                 
 \def\PLambda      {\ensuremath{\Lambda}\xspace}                 
 \def\PSigma      {\ensuremath{\Sigma}\xspace}                 
 \def\POmega      {\ensuremath{\Omega}\xspace}                 
 \def\PUpsilon      {\ensuremath{\Upsilon}\xspace}                 
 
 \def\PB      {\ensuremath{\mathrm{B}}\xspace}                 
                  
 \def\PD      {\ensuremath{\mathrm{D}}\xspace}

 \def\PJ      {\ensuremath{\mathrm{J}}\xspace}                 
 \def\PK      {\ensuremath{\mathrm{K}}\xspace}

 \def\Pb      {\ensuremath{\mathrm{b}}\xspace}                 
 \def\Pc      {\ensuremath{\mathrm{c}}\xspace}                 
 \def\Pd      {\ensuremath{\mathrm{d}}\xspace}

 \def\Pi      {\ensuremath{\mathrm{i}}\xspace}

 \def\Pn      {\ensuremath{\mathrm{n}}\xspace}

 \def\Ps      {\ensuremath{\mathrm{s}}\xspace}                 
 \def\Pt      {\ensuremath{\mathrm{t}}\xspace}

}
{

 \def\Pmu         {\ensuremath{\mu}\xspace}                 
 \def\Pnu         {\ensuremath{\nu}\xspace}                 
                  
 \def\Ppi         {\ensuremath{\pi}\xspace}

 \def\Ppsi        {\ensuremath{\psi}\xspace}                 
                  
 \mathchardef\PDelta="7101
 \mathchardef\PXi="7104
 \mathchardef\PLambda="7103
 \mathchardef\PSigma="7106
 \mathchardef\POmega="710A
 \mathchardef\PUpsilon="7107
                  
 \def\PB      {\ensuremath{B}\xspace}                 
                  
 \def\PD      {\ensuremath{D}\xspace}

 \def\PJ      {\ensuremath{J}\xspace}                 
 \def\PK      {\ensuremath{K}\xspace}

 \def\Pb      {\ensuremath{b}\xspace}                 
 \def\Pc      {\ensuremath{c}\xspace}                 
 \def\Pd      {\ensuremath{d}\xspace}

 \def\Pi      {\ensuremath{i}\xspace}

 \def\Pn      {\ensuremath{n}\xspace}

 \def\Ps      {\ensuremath{s}\xspace}                 
 \def\Pt      {\ensuremath{t}\xspace}

}

\makeatletter
\ifcase \@ptsize \relax
  \newcommand{\miniscule}{\@setfontsize\miniscule{4}{5}}
\or
  \newcommand{\miniscule}{\@setfontsize\miniscule{5}{6}}
\or
  \newcommand{\miniscule}{\@setfontsize\miniscule{5}{6}}
\fi
\makeatother

\DeclareRobustCommand{\optbar}[1]{\shortstack{{\miniscule (\rule[.5ex]{1.25em}{.18mm})}
  \\ [-.7ex] $#1$}}

\def\mup        {{\ensuremath{\Pmu^+}}\xspace}
\def\mun        {{\ensuremath{\Pmu^-}}\xspace} 

\def\neu        {{\ensuremath{\Pnu}}\xspace}
\def\neub       {{\ensuremath{\overline{\Pnu}}}\xspace}
\def\neum       {{\ensuremath{\neu_\mu}}\xspace}
\def\neumb      {{\ensuremath{\neub_\mu}}\xspace}

\def\dquark    {{\ensuremath{\Pd}}\xspace}

\def\squark    {{\ensuremath{\Ps}}\xspace}

\def\cquark    {{\ensuremath{\Pc}}\xspace}

\def\bquark    {{\ensuremath{\Pb}}\xspace}

\def\tquark    {{\ensuremath{\Pt}}\xspace}

\def\pion   {{\ensuremath{\Ppi}}\xspace}

\def\pim    {{\ensuremath{\pion^-}}\xspace}

\def\kaon    {{\ensuremath{\PK}}\xspace}
  \def\Kbar    {{\kern 0.2em\overline{\kern -0.2em \PK}{}}\xspace}

\def\KorKbar    {\kern 0.18em\optbar{\kern -0.18em K}{}\xspace}

\def\Kp      {{\ensuremath{\kaon^+}}\xspace}
\def\Km      {{\ensuremath{\kaon^-}}\xspace}

  \def\Dbar    {{\kern 0.2em\overline{\kern -0.2em \PD}{}}\xspace}
\def\D       {{\ensuremath{\PD}}\xspace}

\def\DorDbar    {\kern 0.18em\optbar{\kern -0.18em D}{}\xspace}
\def\Dz      {{\ensuremath{\D^0}}\xspace}
\def\Dzb     {{\ensuremath{\Dbar{}^0}}\xspace}

\def\Dm      {{\ensuremath{\D^-}}\xspace}

\def\Dstar   {{\ensuremath{\D^*}}\xspace}

\def\Dstarm  {{\ensuremath{\D^{*-}}}\xspace}

\def\Ds      {{\ensuremath{\D^+_\squark}}\xspace}

\def\Dsm     {{\ensuremath{\D^-_\squark}}\xspace}

\def\B       {{\ensuremath{\PB}}\xspace}
\def\Bbar    {{\ensuremath{\kern 0.18em\overline{\kern -0.18em \PB}{}}}\xspace}

\def\BorBbar    {\kern 0.18em\optbar{\kern -0.18em B}{}\xspace}
\def\Bz      {{\ensuremath{\B^0}}\xspace}

\def\Bu      {{\ensuremath{\B^+}}\xspace}

\def\Bp      {{\ensuremath{\Bu}}\xspace}

\def\Bd      {{\ensuremath{\B^0}}\xspace}
\def\Bs      {{\ensuremath{\B^0_\squark}}\xspace}

\def\Bdb     {{\ensuremath{\Bbar{}^0}}\xspace}

\def\jpsi     {{\ensuremath{{\PJ\mskip -3mu/\mskip -2mu\Ppsi\mskip 2mu}}}\xspace}

  \def\Y#1S{\ensuremath{\PUpsilon{(#1S)}}\xspace}

\def\neutron     {{\ensuremath{\Pn}}\xspace}

\def\Lz          {{\ensuremath{\PLambda}}\xspace}
\def\Lbar        {{\ensuremath{\kern 0.1em\overline{\kern -0.1em\PLambda}}}\xspace}
\def\LorLbar    {\kern 0.18em\optbar{\kern -0.18em \PLambda}{}\xspace}

\def\Lb      {{\ensuremath{\Lz^0_\bquark}}\xspace}

\def\Lc      {{\ensuremath{\Lz^+_\cquark}}\xspace}

\newcommand{\decay}[2]{\ensuremath{#1\!\to #2}\xspace}         

\def\to                 {\ensuremath{\rightarrow}\xspace}

\def\CP                {{\ensuremath{C\!P}}\xspace}

\def\Vtb  {{\ensuremath{V_{\tquark\bquark}}}\xspace}

\def\Vtds  {{\ensuremath{V_{\tquark\dquark}^\ast}}\xspace}

\newcommand{\dmd}{{\ensuremath{\Delta m_{\dquark}}}\xspace}

\newcommand{\Gd}{{\ensuremath{\Gamma_{\dquark}}}\xspace}

\newcommand{\mistag}{\ensuremath{\omega}\xspace}

\newcommand{\etag}{{\ensuremath{\varepsilon_{\rm tag}}}\xspace}

\newcommand{\efftag}{{\ensuremath{\etag(1-2\omega)^2}}\xspace}

\def\BuToJPsiK    {\decay{\Bu}{\jpsi\Kp}}

\def\AT#1     {\ensuremath{A_{\mathrm{T}}^{#1}}\xspace}           

\def\C#1      {\ensuremath{\mathcal{C}_{#1}}\xspace}                       
\def\Cp#1     {\ensuremath{\mathcal{C}_{#1}^{'}}\xspace}                    
\def\Ceff#1   {\ensuremath{\mathcal{C}_{#1}^{\mathrm{(eff)}}}\xspace}        
\def\Cpeff#1  {\ensuremath{\mathcal{C}_{#1}^{'\mathrm{(eff)}}}\xspace}       
\def\Ope#1    {\ensuremath{\mathcal{O}_{#1}}\xspace}                       
\def\Opep#1   {\ensuremath{\mathcal{O}_{#1}^{'}}\xspace}                    

\newcommand{\tev}{\ifthenelse{\boolean{inbibliography}}{\ensuremath{~T\kern -0.05em eV}\xspace}{\ensuremath{\mathrm{\,Te\kern -0.1em V}}}\xspace}
\newcommand{\gev}{\ensuremath{\mathrm{\,Ge\kern -0.1em V}}\xspace}
\newcommand{\mev}{\ensuremath{\mathrm{\,Me\kern -0.1em V}}\xspace}
\newcommand{\kev}{\ensuremath{\mathrm{\,ke\kern -0.1em V}}\xspace}
\newcommand{\ev}{\ensuremath{\mathrm{\,e\kern -0.1em V}}\xspace}
\newcommand{\gevc}{\ensuremath{{\mathrm{\,Ge\kern -0.1em V\!/}c}}\xspace}
\newcommand{\mevc}{\ensuremath{{\mathrm{\,Me\kern -0.1em V\!/}c}}\xspace}
\newcommand{\gevcc}{\ensuremath{{\mathrm{\,Ge\kern -0.1em V\!/}c^2}}\xspace}
\newcommand{\gevgevcccc}{\ensuremath{{\mathrm{\,Ge\kern -0.1em V^2\!/}c^4}}\xspace}
\newcommand{\mevcc}{\ensuremath{{\mathrm{\,Me\kern -0.1em V\!/}c^2}}\xspace}

\def\mum  {\ensuremath{{\,\upmu\rm m}}\xspace}

\def\invfb   {\ensuremath{\mbox{\,fb}^{-1}}\xspace}

\def\ps   {\ensuremath{{\rm \,ps}}\xspace}
\def\fs   {\ensuremath{\rm \,fs}\xspace}

\def\invns{\ensuremath{{\rm \,ns^{-1}}}\xspace}

\newcommand{\chisq}{\ensuremath{\chi^2}\xspace}

\def\gsim{{~\raise.15em\hbox{$>$}\kern-.85em
          \lower.35em\hbox{$\sim$}~}\xspace}
\def\lsim{{~\raise.15em\hbox{$<$}\kern-.85em
          \lower.35em\hbox{$\sim$}~}\xspace}

\def\sPlot{\mbox{\em sPlot}}

\def\sqs   {\ensuremath{\protect\sqrt{s}}\xspace}

\def\ptot       {\mbox{$p$}\xspace}
\def\pt         {\mbox{$p_{\rm T}$}\xspace}

\def\evtgen     {\mbox{\textsc{EvtGen}}\xspace}

\def\geant      {\mbox{\textsc{Geant4}}\xspace}

\def\photos     {\mbox{\textsc{Photos}}\xspace}

\def\pythia     {\mbox{\textsc{Pythia}}\xspace}

\def\tell1  {TELL1\xspace}
\def\ukl1   {UKL1\xspace}

\def\BdToDstMuNu    {\decay{\Bd}{\Dstarm\mup\neum X}}
\def\BdToDmMuNu     {\decay{\Bd}{\Dm\mup\neum X}}
\def\DstarMode      {\BdToDstMuNu}
\def\DstMode        {\DstarMode}
\def\DmMode         {\BdToDmMuNu}
\def\DpMode         {\DmMode}

\def\BuToDstMuNu    {\decay{\Bu}{\Dstarm\mup\neum X}}

\def\BuDstMode      {\BuToDstMuNu}

\def\DmToKpipi      {\decay{\Dm}{\Kp\pim\pim}}
\def\DstarmToDpi    {\decay{\Dstarm}{\Dzb\pim}}
\def\DzbToKpi       {\decay{\Dzb}{\Kp\pim}}

\def\DstDm          {\ensuremath{\D^{(*)-}}\xspace}
\def\DstDp          {\ensuremath{\D^{(*)+}}\xspace}
\def\BdToDorDstMuNu    {\decay{\Bd}{\DstDm\mup\neum X}}
\def\BuToDorDstMuNu    {\decay{\Bu}{\DstDm\mup\neum X}}
\def\DorDstarMode {\BdToDorDstMuNu}
\def\BdToDbarorDstarbarMuNu    {\decay{\Bdb}{\DstDp\mun\neumb X}}
\def\DbarorDstarbarMode {\BdToDbarorDstarbarMuNu}

\usepackage{cite} 
\usepackage{mciteplus}

\usepackage{longtable} 

\begin{document}

\renewcommand{\thefootnote}{\fnsymbol{footnote}}
\setcounter{footnote}{1}


\begin{titlepage}
\pagenumbering{roman}

\vspace*{-1.5cm}
\centerline{\large EUROPEAN ORGANIZATION FOR NUCLEAR RESEARCH (CERN)}
\vspace*{1.5cm}
\noindent
\begin{tabular*}{\linewidth}{lc@{\extracolsep{\fill}}r@{\extracolsep{0pt}}}
\ifthenelse{\boolean{pdflatex}}
{\vspace*{-2.7cm}\mbox{\!\!\!\includegraphics[width=.14\textwidth]{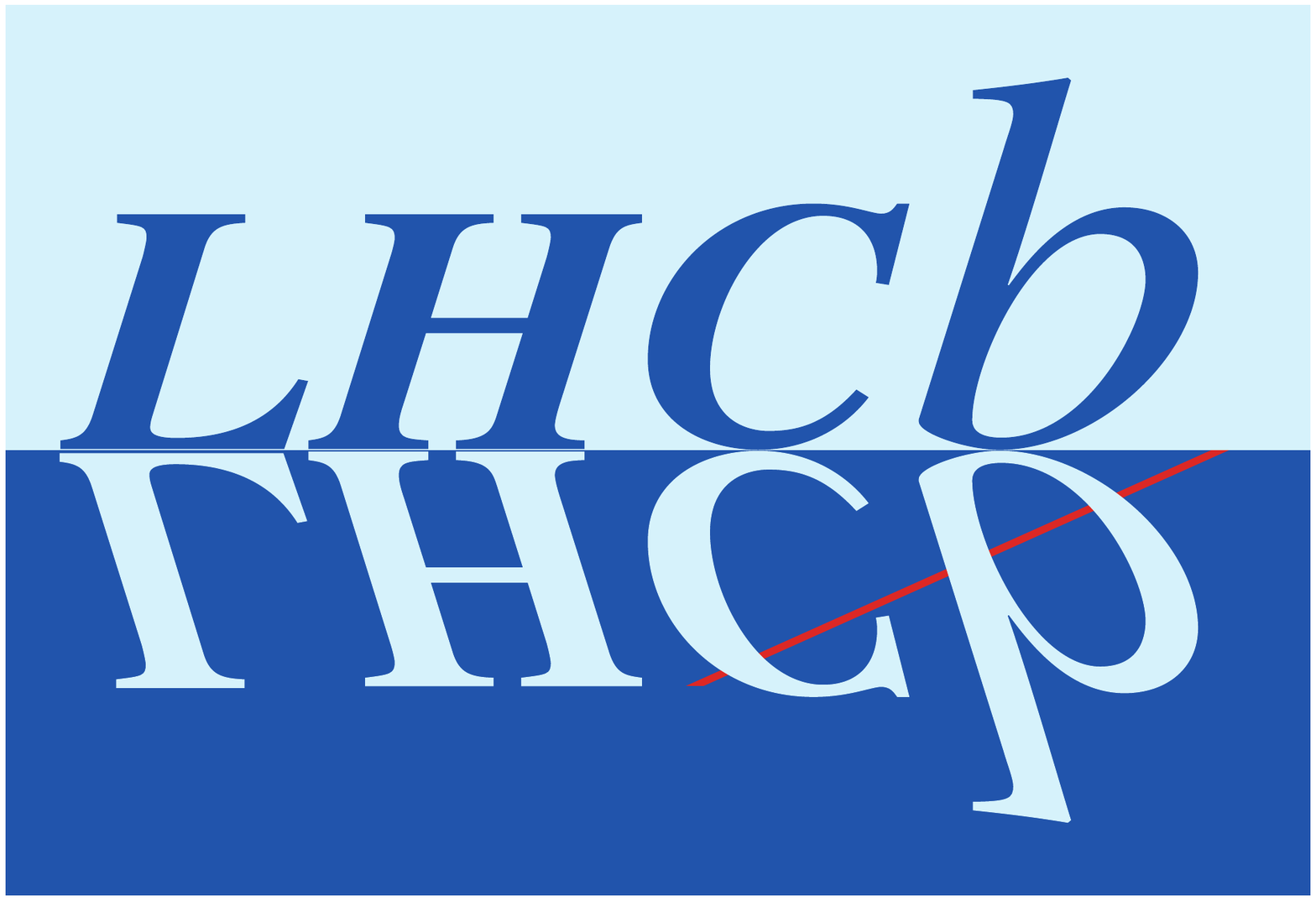}} & &}%
{\vspace*{-1.2cm}\mbox{\!\!\!\includegraphics[width=.12\textwidth]{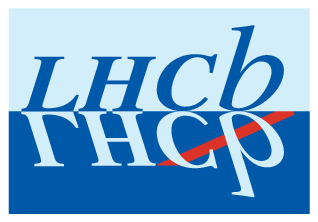}} & &}%
\\
 & & CERN-EP-2016-084 \\  
 & & LHCb-PAPER-2015-031 \\  
 & & 21 July 2016 \\ 
 & & \\
\end{tabular*}

\vspace*{4.0cm}

{\bf\boldmath\huge
\begin{center}
  A precise measurement of the $B^0$ meson oscillation frequency
\end{center}
}

\vspace*{2.0cm}

\begin{center}
The LHCb collaboration\footnote{Authors are listed at the end of this paper.}
\end{center}

\vspace{\fill}

\begin{abstract}
  \noindent
  The oscillation frequency, $\Delta m_d$, of $B^0$ mesons is measured using 
  semileptonic decays with a $D^-$ or $D^{*-}$ meson in the final state. The data 
  sample corresponds to 3.0$\mbox{\,fb}^{-1}$ of $pp$ collisions,
  collected by the LHCb experiment at centre-of-mass energies
  $\sqrt{s}$ = 7 and 8$\rm \,Te\kern -0.1em V$. A combination of the two decay modes gives 
  $\Delta m_d = (505.0 \pm 2.1 \pm 1.0) \rm \,ns^{-1}$, 
  where the first uncertainty is statistical and the second is systematic. 
  This is the most precise single measurement of this parameter. 
  It is consistent with the current world average and has similar precision.
\end{abstract}

\vspace*{2.0cm}

\begin{center}
Published in Eur. Phys. J. C (2016) 76: 412.
\end{center}

\vspace{\fill}

{\footnotesize 
\centerline{\copyright~CERN on behalf of the \lhcb collaboration, licence \href{http://creativecommons.org/licenses/by/4.0/}{CC-BY-4.0}.}}
\vspace*{2mm}

\end{titlepage}


\newpage
\setcounter{page}{2}
\mbox{~}

\cleardoublepage


\renewcommand{\thefootnote}{\arabic{footnote}}
\setcounter{footnote}{0}



\pagestyle{plain} 
\setcounter{page}{1}
\pagenumbering{arabic}


\section{Introduction}
\label{sec:Introduction}

Flavour oscillation, or mixing, of neutral meson systems gives mass eigenstates that are different from flavour eigenstates. 
In the \Bd--\Bdb system, the mass difference between mass eigenstates, \dmd, is 
directly related to the square of the product of the CKM matrix
elements $\Vtb$ and $\Vtds$, and is therefore sensitive to fundamental parameters 
of the Standard Model, as well as to non-perturbative
strong-interaction effects and the square of the top quark mass~\cite{MixPDG2014}. 
Measurements of mixing of neutral \B mesons were published for the first time by UA1~\cite{Albajar:1986it} and ARGUS~\cite{Prentice:1987ap}. 
Measurements of \Bd--\Bdb mixing  have been performed by
CLEO~\cite{Behrens:2000qu}, 
experiments at LEP and SLC~\cite{Abbaneo:2000ej}, experiments at the Tevatron~\cite{Abazov:2006qp,Affolder:1999cn}, 
the \B Factories experiments~\cite{Aubert:2005kf,Abe:2004mz} and, most recently, at
LHCb~\cite{LHCb-PAPER-2011-010,LHCb-PAPER-2013-036,Aaij:2012nt}. 
The combined world average value for the mass difference, $\dmd=(510
\pm 3)$\invns, has a relative precision of 0.6\%~\cite{PDG2014}. 
This paper reports a measurement of \dmd based on \DmMode and
\DstMode decays,\footnote{The inclusion of charge-conjugate processes
  is implied throughout.} 
where $X$ indicates any additional particles that are not
reconstructed. The data sample used for this measurement was collected
at LHCb during LHC Run 1 at {\sqs=\,7\,(8)\tev} in 2011\,(2012), corresponding to integrated
luminosities of $1.0$ ($2.0$)\invfb.

The relatively high branching fraction for semileptonic decays of
\Bd mesons, along with the highly efficient lepton
identification and flavour tagging
capabilities at \lhcb, results in abundant samples of \DorDstarMode
decays, where the flavour of the \Bd meson at the time of production
and decay can be inferred. 
In addition, the decay time $t$ of \Bd mesons can be determined with 
adequate resolution, even though the decay is not fully reconstructed,
because of the potential presence of undetected particles. 
It is therefore possible to precisely measure \dmd as the 
frequency of matter-antimatter oscillations in a time-dependent
analysis of the decay rates of unmixed and mixed events, 
\begin{align}
  N^{\rm unmix}(t) \equiv N(\DorDstarMode)(t) \propto e^{-\Gd t} [1 + \cos (\dmd t)] \ ,\nonumber \\
  N^{\rm mix}(t)  \equiv N(\Bz\to \DbarorDstarbarMode)(t) \propto
  e^{-\Gd t} [ 1 - \cos (\dmd t) ] \ , 
\label{eqn:mixunmix}
\end{align} 
\noindent where the state assignment is based on the flavours of the \Bd meson at production and
decay, which may be the same (unmixed) or opposite (mixed).  
In Eqn.~\ref{eqn:mixunmix}, 
$\Gd=1/\tau_{\Bz}$ is the decay width of the \Bd meson, $\tau_{\Bz}$
being its lifetime. Also, in Eqn.~\ref{eqn:mixunmix} the difference in the decay widths
of the mass eigenstates, $ \Delta \Gd, $ and
\CP\ violation in mixing are neglected, due to their negligible impact
on the results. 
The flavour asymmetry between unmixed and mixed events is
\begin{equation}
A(t) = \frac{N^{\rm unmix}(t) - N^{\rm mix} (t)}{N^{\rm unmix}(t) +
  N^{\rm mix}(t)} = \cos (\dmd t)  \ .
\end{equation}

A description of the \lhcb detector and the 
datasets used in this measurement is given in Sec.~\ref{sec:Detector}. 
Section~\ref{sec:eventSelection} presents the selection criteria, the flavour tagging
algorithms, and the method chosen to reconstruct the \Bd decay time. 
The fitting strategy and results are described in
Sec.~\ref{sec:fitting}. 
A summary of the systematic uncertainties is given in 
Sec.~\ref{sec:systematics}, and conclusions are reported in Sec.~\ref{sec:conclusion}.

\section{Detector and simulation}
\label{sec:Detector}
The \lhcb detector~\cite{Alves:2008zz,LHCb-DP-2014-002} 
is a single-arm forward spectrometer covering the \mbox{pseudorapidity} range $2<\eta <5$,
designed for the study of particles containing \bquark or \cquark
quarks. The detector includes a high-precision tracking system
consisting of a silicon-strip vertex detector surrounding the $pp$
interaction region, 
a large-area silicon-strip detector located
upstream of a dipole magnet with a bending power of about
$4{\rm\,Tm}$, and three stations of silicon-strip detectors and straw
drift tubes 
placed downstream of the magnet.
The tracking system provides a measurement of momentum, \ptot, of charged particles with
a relative uncertainty that varies from 0.5\% at low momentum to 1.0\% at 200\gevc.
The minimum distance of a track to a primary vertex (PV), the impact parameter (IP), is measured with a resolution of $(15+29/\pt)\mum$,
where \pt is the component of the momentum transverse to the beam, in\,\gevc.
Different types of charged hadrons are distinguished using information
from two ring-imaging Cherenkov (RICH) detectors. 
Photons, electrons and hadrons are identified by a calorimeter system consisting of
scintillating-pad and preshower detectors, an electromagnetic
calorimeter and a hadronic calorimeter. Muons are identified by a
system composed of alternating layers of iron and multiwire
proportional chambers.

The online event selection is performed by a
trigger~\cite{LHCb-PUB-2014-046}, 
which consists of a hardware stage, based on information from the calorimeter and muon
systems, followed by a software stage, which applies a full event
reconstruction.
Candidate events are first required to pass the hardware trigger,
  which selects muons with a transverse momentum $\pt>1.48\gevc$
  in the 7\tev data or $\pt>1.76\gevc$ in the 8\tev data. 
The software trigger requires a two-, three- or four-track
secondary vertex, where one of the tracks is identified as a muon, 
with a significant displacement from the primary
$pp$ interaction vertices. At least one charged particle
must have a transverse momentum $\pt > 1.7\gevc$ and be
inconsistent with originating from a PV. As it will be explained
later, the software trigger selection introduces a bias on the \dmd 
measurement, which is corrected for.
A multivariate algorithm~\cite{BBDT} is used for
the identification of secondary vertices consistent with the decay
of a \bquark hadron.

The method chosen to reconstruct the \Bd decay time relies on Monte Carlo simulation. 
Simulation is also used to estimate the main background sources and to verify the fit model. 
In the simulation, $pp$ collisions are generated using
\pythia~\cite{Sjostrand:2006za,*Sjostrand:2007gs} 
 with a specific \lhcb
configuration~\cite{LHCb-PROC-2010-056}.  Decays of hadronic particles
are described by \evtgen~\cite{Lange:2001uf}, in which final-state
radiation is generated using \photos~\cite{Golonka:2005pn}. The
interaction of the generated particles with the detector, and its response,
are implemented using the \geant
toolkit~\cite{Allison:2006ve, *Agostinelli:2002hh} as described in
Ref.~\cite{LHCb-PROC-2011-006}.
Large samples of mixtures of semileptonic decays resulting in a \Dm or a \Dstarm meson in the final state were 
simulated and the assumptions used to build these samples are
assessed in the evaluation of systematic uncertainties.

\section{Event selection}
\label{sec:eventSelection}

For charged particles used to reconstruct signal candidates,
requirements 
are imposed on track quality, momentum, transverse momentum, and impact parameter with
respect to any PV.  
Tracks are required to be identified as muons, kaons or pions.
The charm mesons are reconstructed through the \DmToKpipi decay, or
through the \DstarmToDpi, \DzbToKpi decay chain. 
The masses of the reconstructed \Dm and \Dzb mesons
should be 
within $70\mevcc$ and $40\mevcc$ of their known values~\cite{PDG2014}, 
while the mass difference between the
reconstructed \Dstarm and \Dzb mesons should lie between $140\mevcc$ and
$155\mevcc$.  
For \Dm and \Dzb candidates, the scalar sum of the \pt of the daughter
tracks should be
above $1800\mevc$. 
A good quality vertex fit is required
for the \Dm, \Dzb, and \Dstarm candidates, and for the
$D^{(*)-}\mu^+$ combinations. When more than one combination is found
in an event,
the one with the smallest vertex \chisq (hereafter referred to as the \PB
candidate) is chosen. 
The reconstructed vertices of \Dm, \Dzb, and \PB candidates are required
to be significantly displaced from their associated PV, 
where the associated PV is that which has the smallest \chisq increase when adding the candidate.
For \Dm and \Dzb candidates, a large IP with respect to the associated PV is required
in order to suppress charm mesons promptly produced in $pp$ collisions. 
The momentum of the \PB candidate, and its flight direction measured using the PV and the \PB
vertex positions, are required to be aligned. 
These selection criteria reduce  to the per-mille level or lower the contribution of
\DstDm decays where the charmed meson originates from the PV.
The invariant mass of the \PB candidate is required to be  in the range $[3.0,5.2]\gevcc$. 

Backgrounds from $\PB \to \jpsi X$ decays, where one of the muons from
the $\jpsi \to \mup\mun$ decay is correctly identified and the other misidentified as a pion and used to reconstruct a \DstDm, 
are suppressed by applying a veto around the $\jpsi$ mass. 
Similarly, a veto around the \Lc mass is applied to suppress semileptonic decays of the \Lb baryon, 
in which the proton of the subsequent \Lc decay into $p K^- \pi^+$ is misidentified as a pion.

The dominant background is due to \BuToDorDstMuNu decays, where additional particles coming from the decay of higher
charm resonances, or from multi-body decays of \Bu mesons, are
neglected. 
The fractions of \Bu decays in the \Dm and \Dstarm samples are expected
to be 13\% and 10\%, based on the branching fractions of signal and
background, with uncertainties at the 10\% level. 
This background is reduced by using a multivariate discriminant based on a boosted decision tree~(BDT) algorithm~\cite{Breiman,AdaBoost}, which exploits information on the \PB candidate, 
kinematics of the higher charm resonances and isolation criteria for tracks and composite
candidates in the \PB decay chain. Training of the BDT classifier is carried out using simulation samples of \DstarMode signal and \BuDstMode background.
The variables used as input for the BDT classifier are described in
the Appendix. 
Only candidates with BDT output larger than
$-0.12$ ($-0.16$) are selected in the 2011 (2012) data sample for the
\DpMode mode. The BDT output
is required to be larger than $-0.3$ in both
2011 and 2012 data samples for the \DstarMode
mode. The impact of this requirement on signal efficiency and
background retention can be seen in
Fig.~\ref{fig:IsoBDT_Fit_2012}. The 
background from \Bu decays is reduced by 70\% in both modes. 
Combinatorial background is evaluated by using reconstructed candidates
in the \DstDm signal mass sidebands. 
Backgrounds due to decays of \Bs and \Lb into similar final states to 
those of the signal are studied through  simulations. 

The decay time of the \Bd meson is calculated as 
$t = {(M_{\Bd} \cdot L)}/{(p_{\rm rec}\cdot c / k)}$, 
where $M_{\Bd}$ is the mass of the \Bd, taken from Ref.~\cite{PDG2014}, $L$
is the measured decay length and $p_{\rm rec}$ is the magnitude of the visible momentum, measured from the \DstDm meson and the
muon. 
The correction factor $k$ is determined from simulation by dividing
the visible \Bd  momentum by its true value and taking the average, $k = \langle {p_{\rm rec}}/{p_{\rm true}} \rangle $. 
This correction 
represents the dominant source of uncertainty in the determination of the decay time of the \Bd meson for  $t > 1.5 \ps$. 
Since the $k$-factor depends strongly on the decay kinematics, it is
parametrised by a fourth-order polynomial as a function of the visible
mass of the \Bd candidate as explained in the Appendix. 

The \Bz flavour at production is determined by using information
from the other \bquark hadron present in the event. The decision of flavour tagging 
algorithms~\cite{LHCb-PAPER-2011-027} based on the charge of leptons, kaons and of an inclusively
reconstructed detached vertex, is used for the \DstarMode channel. In the \DmMode channel, which
is subject to a larger \Bp background contamination, the decision of the tagging 
algorithm based on the detached vertex is excluded in order to avoid spurious background asymmetries. 
The statistical uncertainty on \dmd decreases as ${\cal{T}}^{-1/2}$ where
the tagging power is defined as ${\cal{T}}=\efftag$, where 
\etag is the tagging efficiency and \mistag is the
mistag rate. 
To increase the statistical precision, the events are grouped into
four tagging categories of increasing predicted mistag probability $\eta$, 
defined by $\eta \in [0,0.25]$, $[0.25,0.33]$,
$[0.33, 0.41]$, $[0.41, 0.47]$.
The mistag probability $\eta$ is evaluated for each \PB candidate from event and taggers
properties and was calibrated on data using control samples~\cite{LHCb-PAPER-2011-027}. 
The average mistag rates for 
signal and background are taken as free parameters when fitting for \dmd. 
The combined tagging power~\cite{LHCb-PAPER-2011-027} for the \DmMode mode is $(2.38\pm 0.05)$\% and 
$(2.46\pm 0.04)$\% in 2011 and 2012. For the \DstarMode mode, 
the tagging power in 2011 and 2012 is $(2.55\pm 0.07)$\% and 
$(2.32\pm 0.04)$\%.

\section{Fit strategy and results}
\label{sec:fitting}

The fit proceeds as follows. 
First, \DstDm mesons originating from semileptonic \Bd or \Bu decays are separated from the
background coming from combinations of tracks not associated to a
charm meson decay, 
by a fit to the invariant mass distributions of the selected candidates. 
This fit assigns to each event a covariance-weighted quantity
{\em{sWeight}}, which is used in the subsequent 
fits to subtract statistically the contribution of the background by means of the \sPlot\ procedure~\cite{Pivk:2004ty}. 
Then, the contribution of \DstDm from \Bu decays is determined in a
fit to the distributions of the BDT classifier output weighted by signal {\em{sWeights}}. 
Next, a cut is
applied on the BDT output in order to suppress the \Bu background, 
the mass distributions are fitted again, and new {\em{sWeights}} are
determined. 
Finally, the oscillation frequency \dmd is determined by a fit to the
decay time distribution of unmixed and mixed candidates, weighted 
for the signal {\em{sWeights}} determined in the previous step. 

An extended binned maximum likelihood fit to the data distributions
is performed for each stage, simultaneously for the four tagging
categories defined above. 
Data samples collected in 2011
and 2012 are treated separately. 

\begin{figure}[bt]
\begin{center} 
\includegraphics[width=0.49\textwidth]{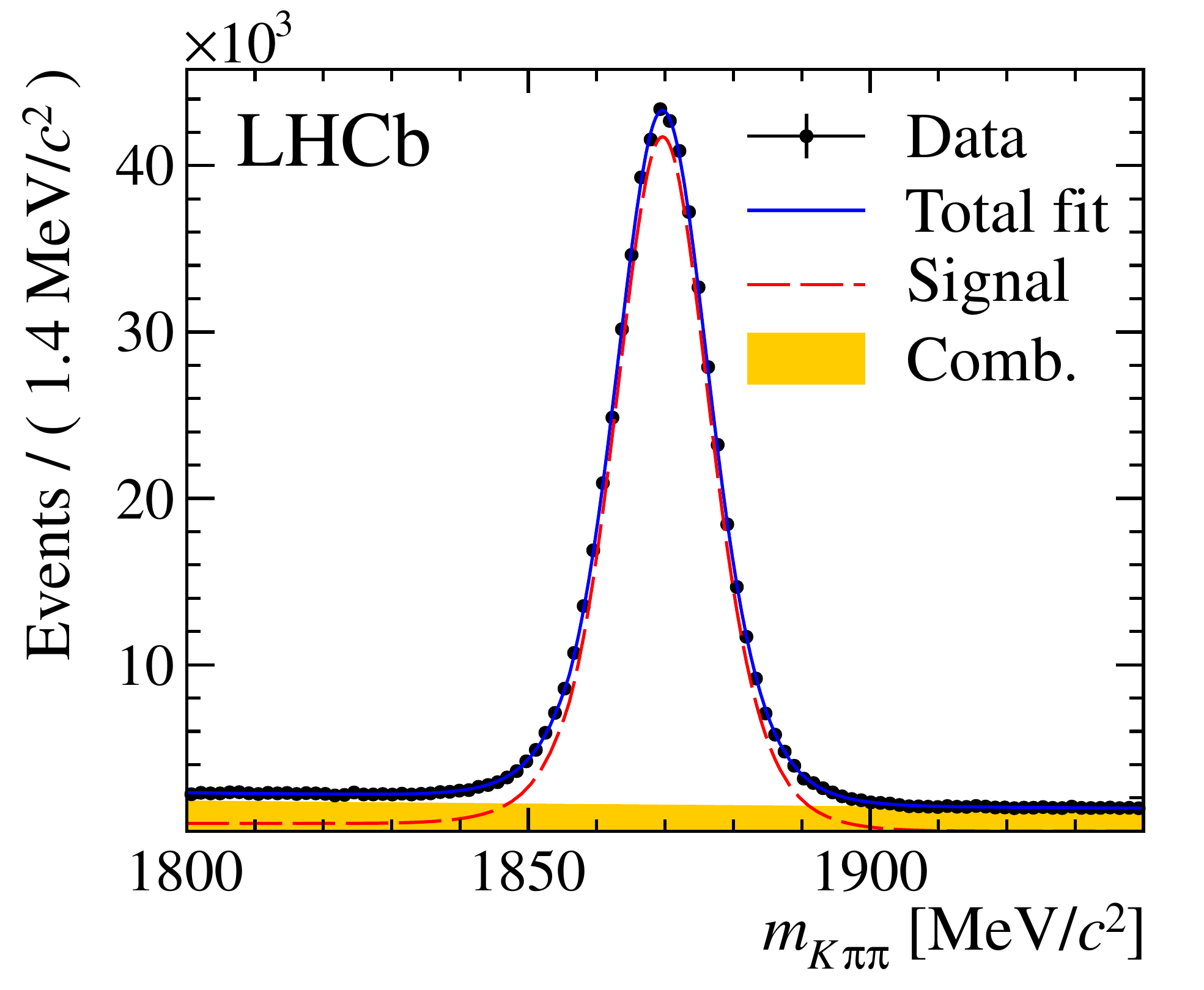}
\includegraphics[width=0.49\textwidth]{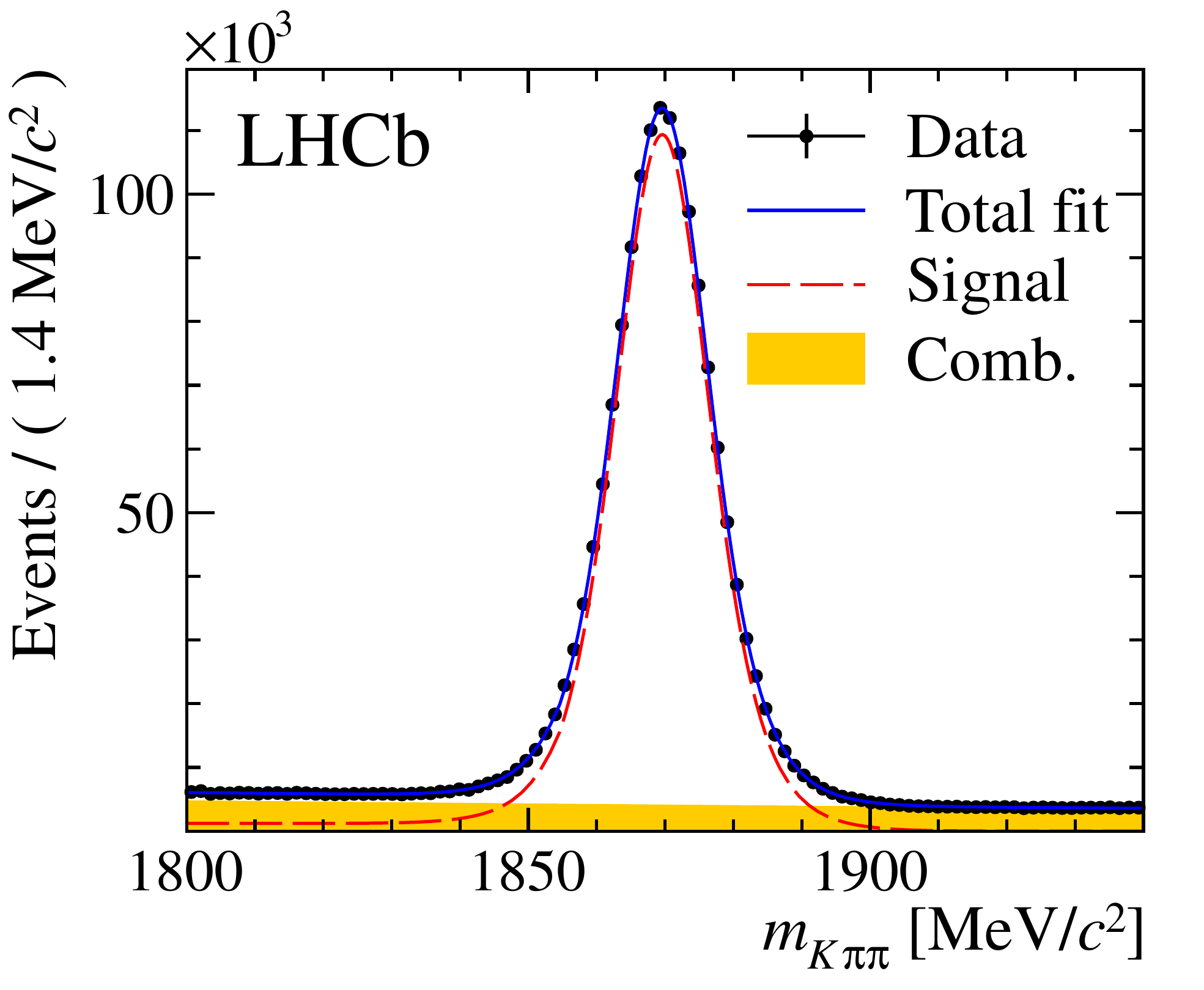}
\caption{\small{Distribution of $m_{K\pi\pi}$ for the \DmMode
    candidates in (left) 2011 and (right) 2012 data. Projections of the 
fit function are superimposed (blue continuous line) for the full PDF 
and its components: (red dashed line) signal \Dm from \Bd or \Bu decays 
and (filled yellow area) combinatorial background. }}
\label{fig:DpMode_MassFit} 
\end{center} 
\end{figure}

Figure~\ref{fig:DpMode_MassFit} shows the results of the fits to the
\Dm candidate mass distributions for \DpMode candidates.   
In these fits, the distributions of \Dm from \Bz and \Bu decays are summed as they are 
described by the same probability density function (PDF): the sum of
two Gaussian functions 
and a Crystal Ball function~\cite{Skwarnicki:1986xj}. 
The yields corresponding to the \Dm peak are $(5.30 \pm 0.02)\times 10^5$ and  
$(1.393 \pm 0.003)\times 10^6$ in 2011 and 2012 data, respectively. 
The combinatorial background, which contributes typically 6\% under the \Dm
peak, is modelled with an exponential distribution. 

\begin{figure}[htb]
\begin{center} 
\includegraphics[width=0.49\textwidth]{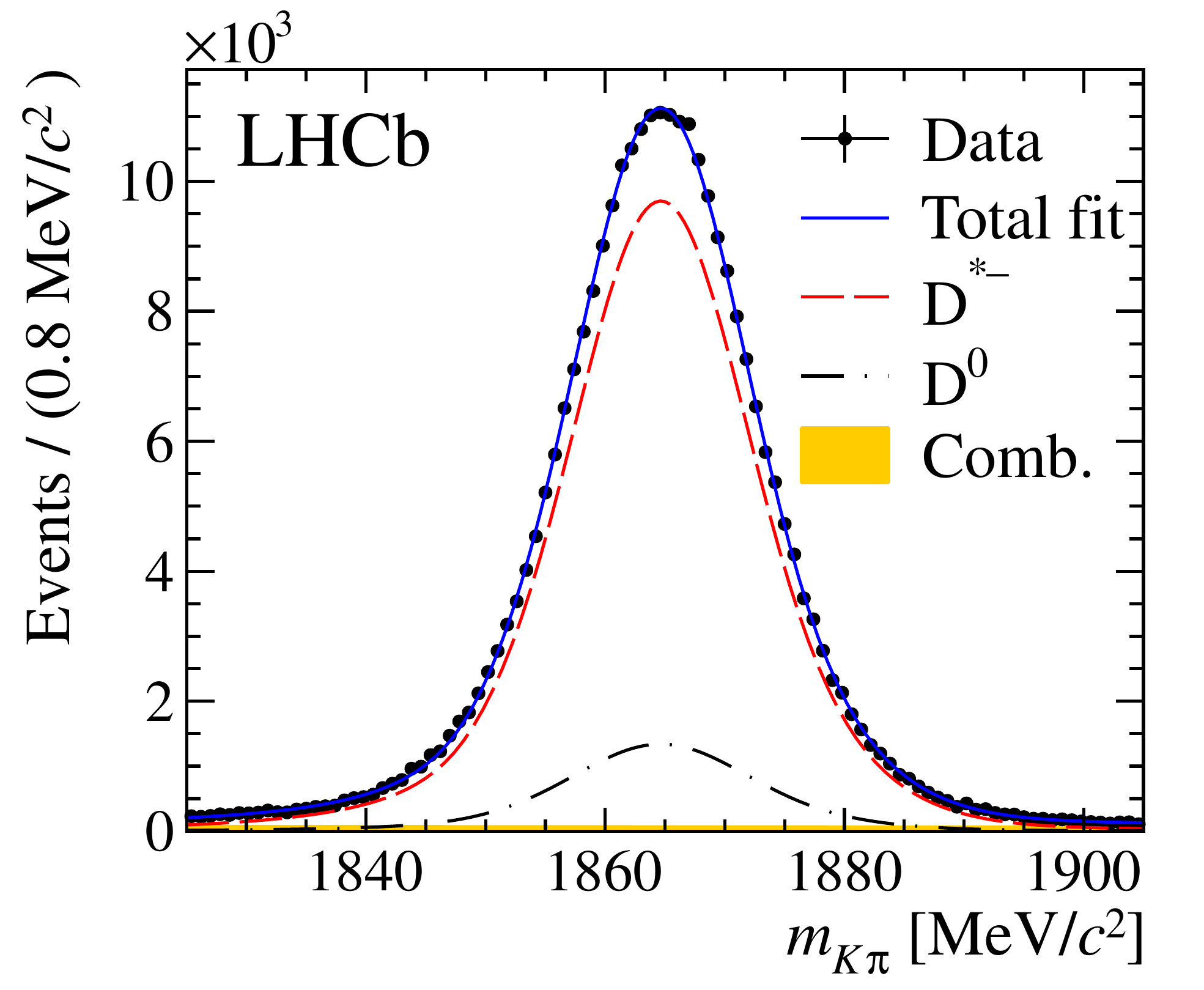}
\includegraphics[width=0.49\textwidth]{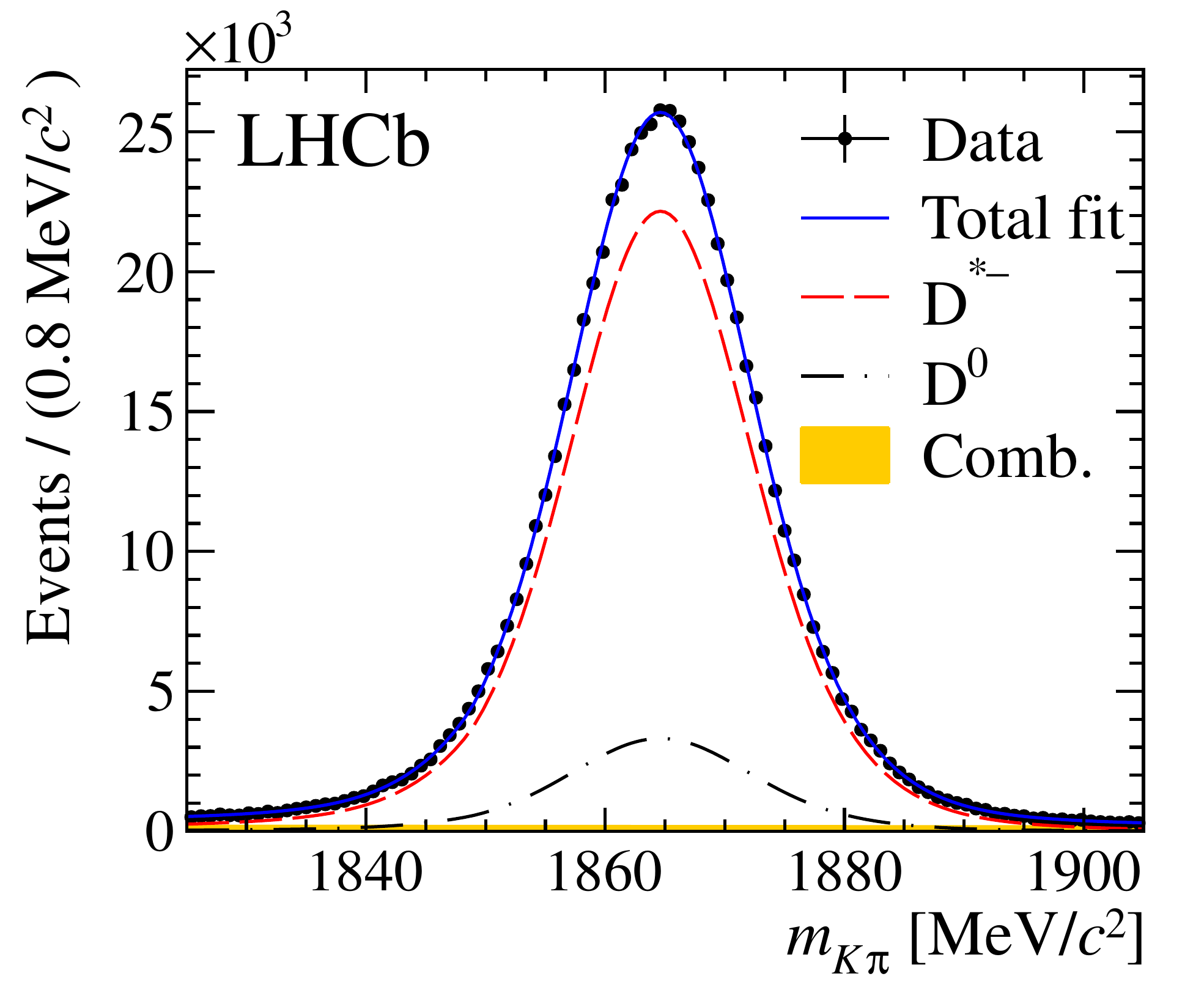}\\
\includegraphics[width=0.49\textwidth]{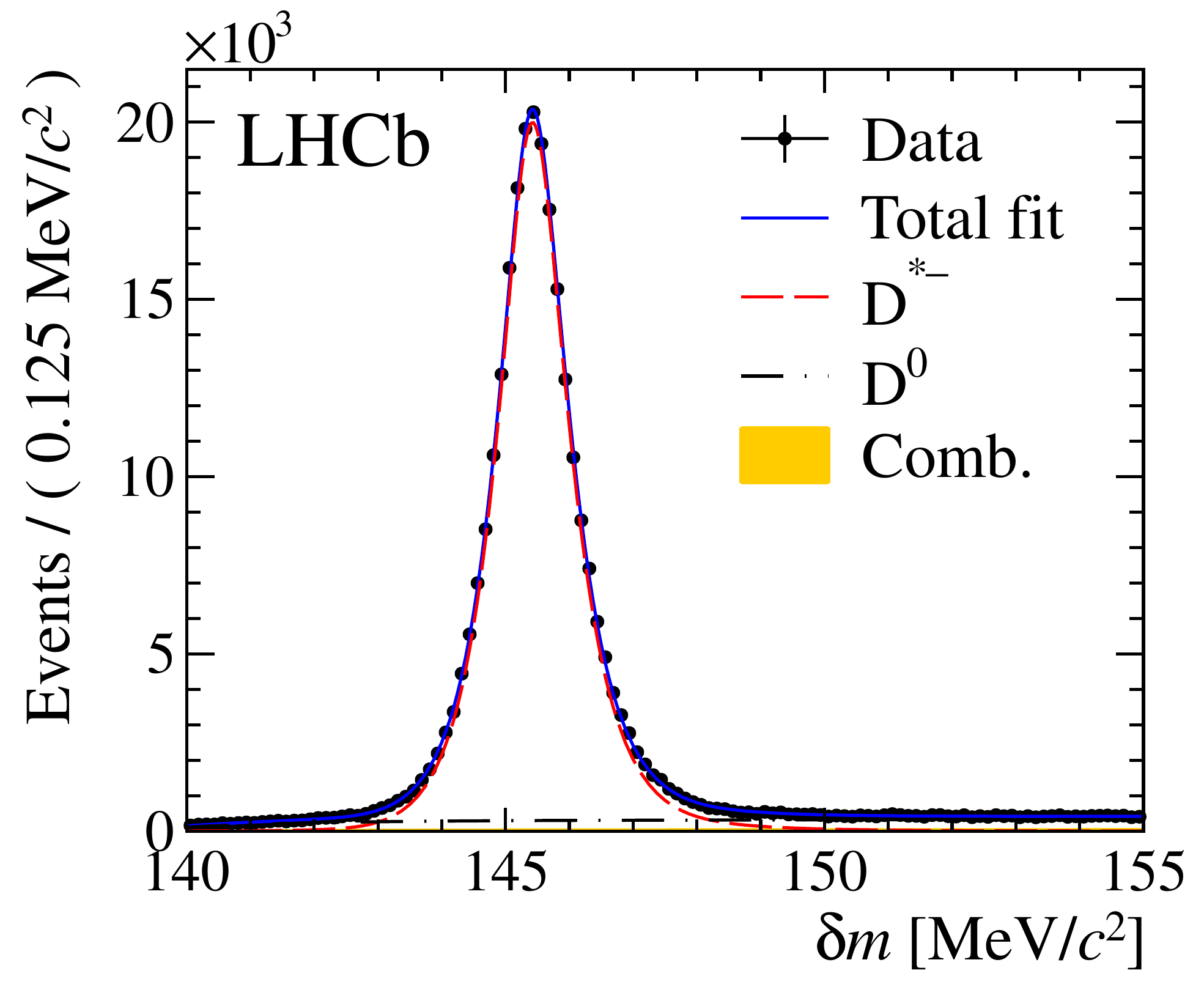}
\includegraphics[width=0.49\textwidth]{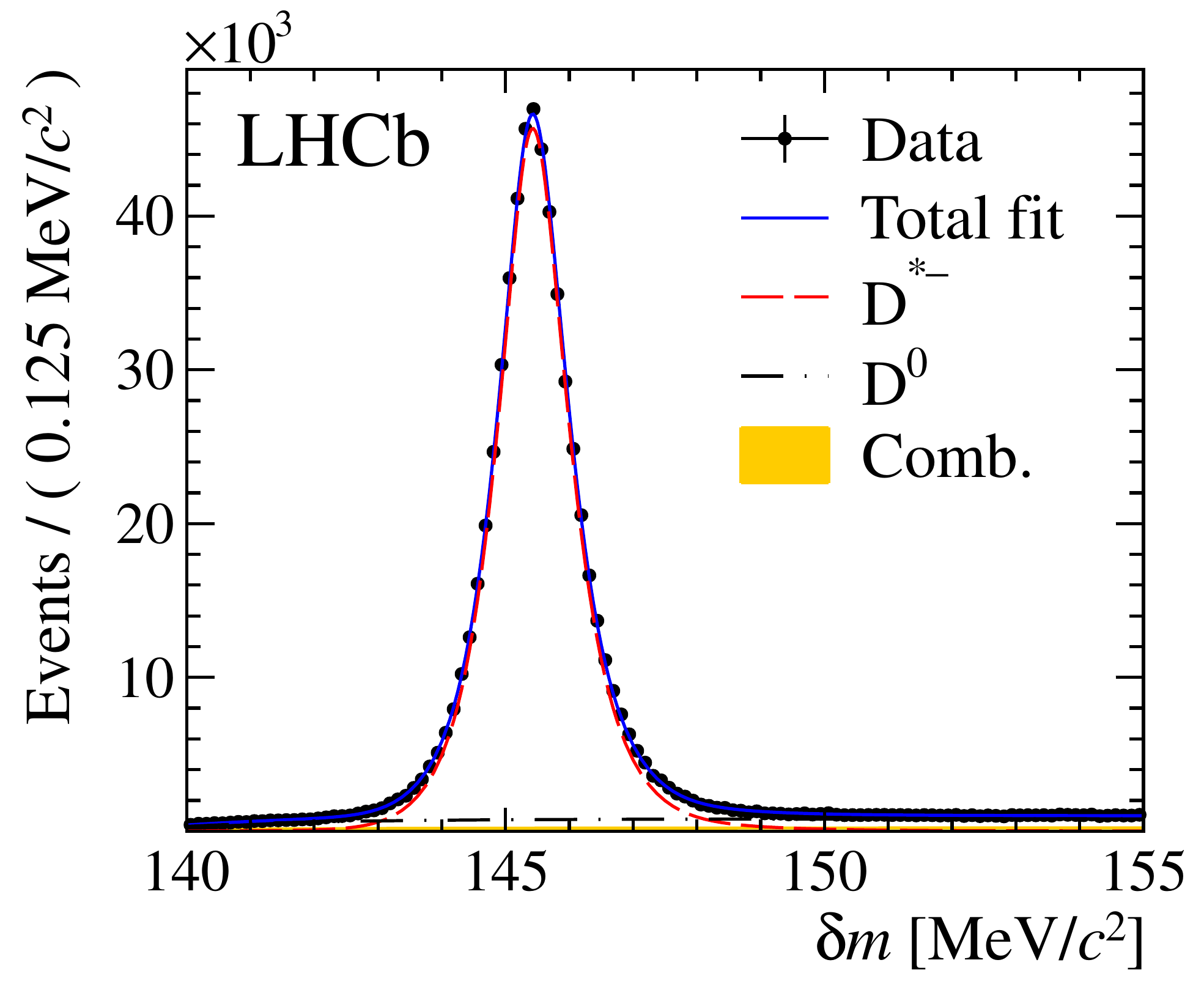}
\caption{\small{Distributions of (top) $m_{K\pi}$ and (bottom) $\delta
    m$ for \DstarMode candidates in (left) 2011 and (right) 2012
    data. Projections of the fit function are superimposed for (blue
    continuous line) the full PDF 
    and its components: (red dashed line) signal \Dstarm from \Bd or \Bu decays,
    (black dashed-dotted line) \Dzb
    from \B and (filled yellow area) combinatorial 
    backgrounds. }
}
\label{fig:DstarMode_MassFit} 
\end{center} 
\end{figure}

For the \DstarMode samples, a simultaneous fit to the distributions of 
the $K^+\pi^-$ invariant mass, $m_{K^+\pi^-}$, and  the invariant mass difference of ${K^+\pi^-\pi^-}$ and  ${K^+\pi^-}$ combinations,
$\delta m = m_{K^+\pi^-\pi^-} - m_{K^+\pi^-}$, is performed.  
Three different components are considered: the signal \Dstar from \Bd
or \Bu decays and two background sources.
The PDF for the mass distributions of \Dstar from \B decays is defined
by the sum of two Gaussian functions  
and a Crystal Ball function in the $m_{K^+\pi^-}$ mass projection and by two 
Gaussian functions and a Johnson function~\cite{JohnsonPDF} in the $\delta m$ mass projection. 
Background candidates containing a \Dzb originating from a \bquark hadron decay 
without an intermediate $D^*$ resonance, which contribute about
15\% in the full $\delta m$ mass range, are described by the same distribution as that of the signal 
for $m_{K^+\pi^-}$, and by an empirical function based on a phase-space
distribution for $\delta m$. 
A combinatorial background
component which contributes typically 0.8\% under the \Dstar peak 
is modelled with an exponential distribution for $m_{K^+\pi^-}$
and the same empirical distribution for $\delta m$ as used for the
\Dzb background. 
All parameters that describe signal and background shapes are allowed to vary freely in the invariant mass fits. 
The results of the 2011 and 2012 fits for these parameters are compatible within the
statistical uncertainties.  Figure~\ref{fig:DstarMode_MassFit} shows
the results of the fit to the \DstarMode samples, projected onto the two
mass observables. 
The yields corresponding to the \Dstar peak 
are $(2.514 \pm 0.006) \times 10^5$ and $(5.776 \pm 0.009)\times 10^5$ in 2011 and 2012 data.

\begin{figure}
  \begin{center} 
   \includegraphics[width=0.85\textwidth]{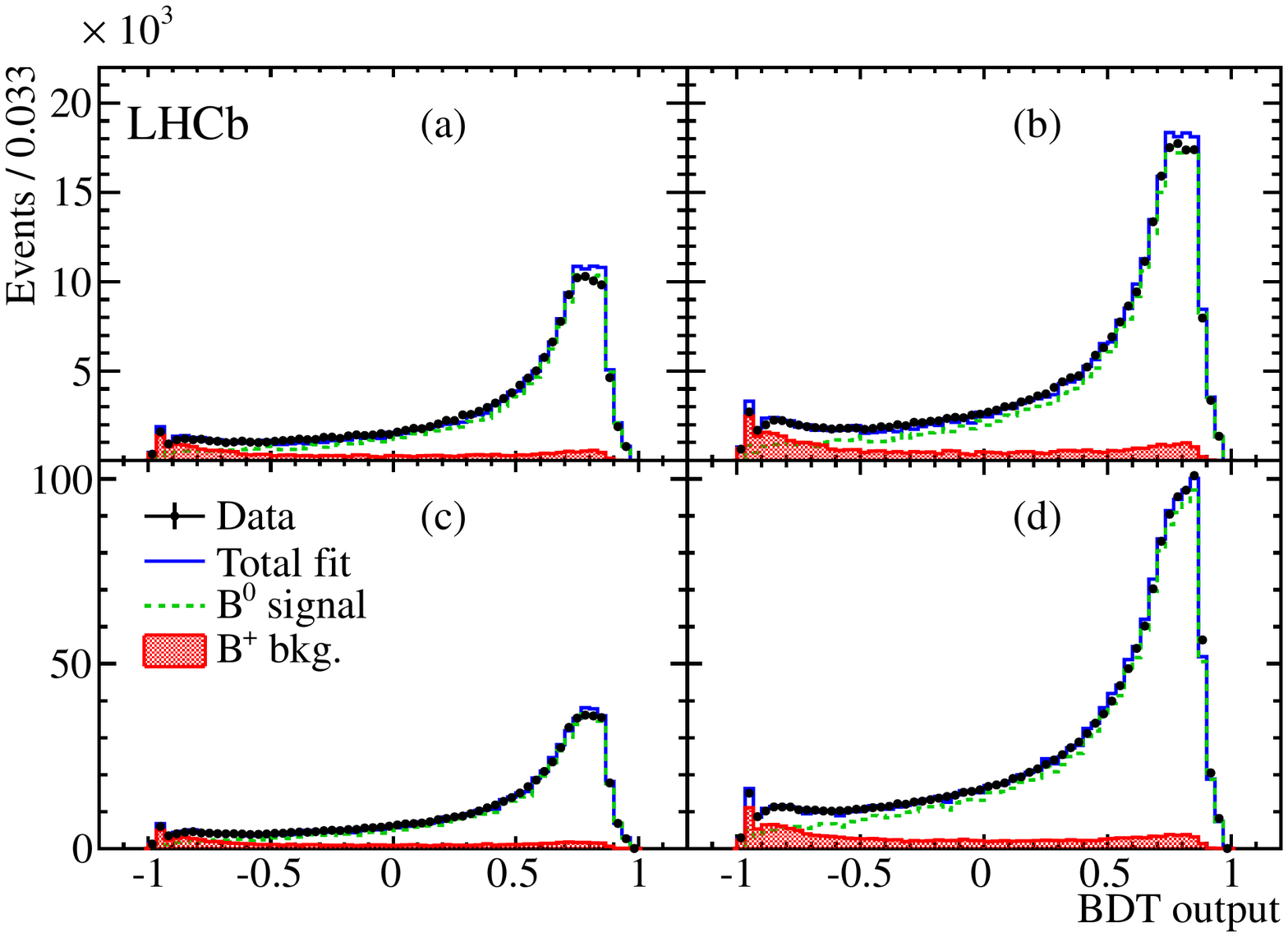}\\
   \includegraphics[width=0.85\textwidth]{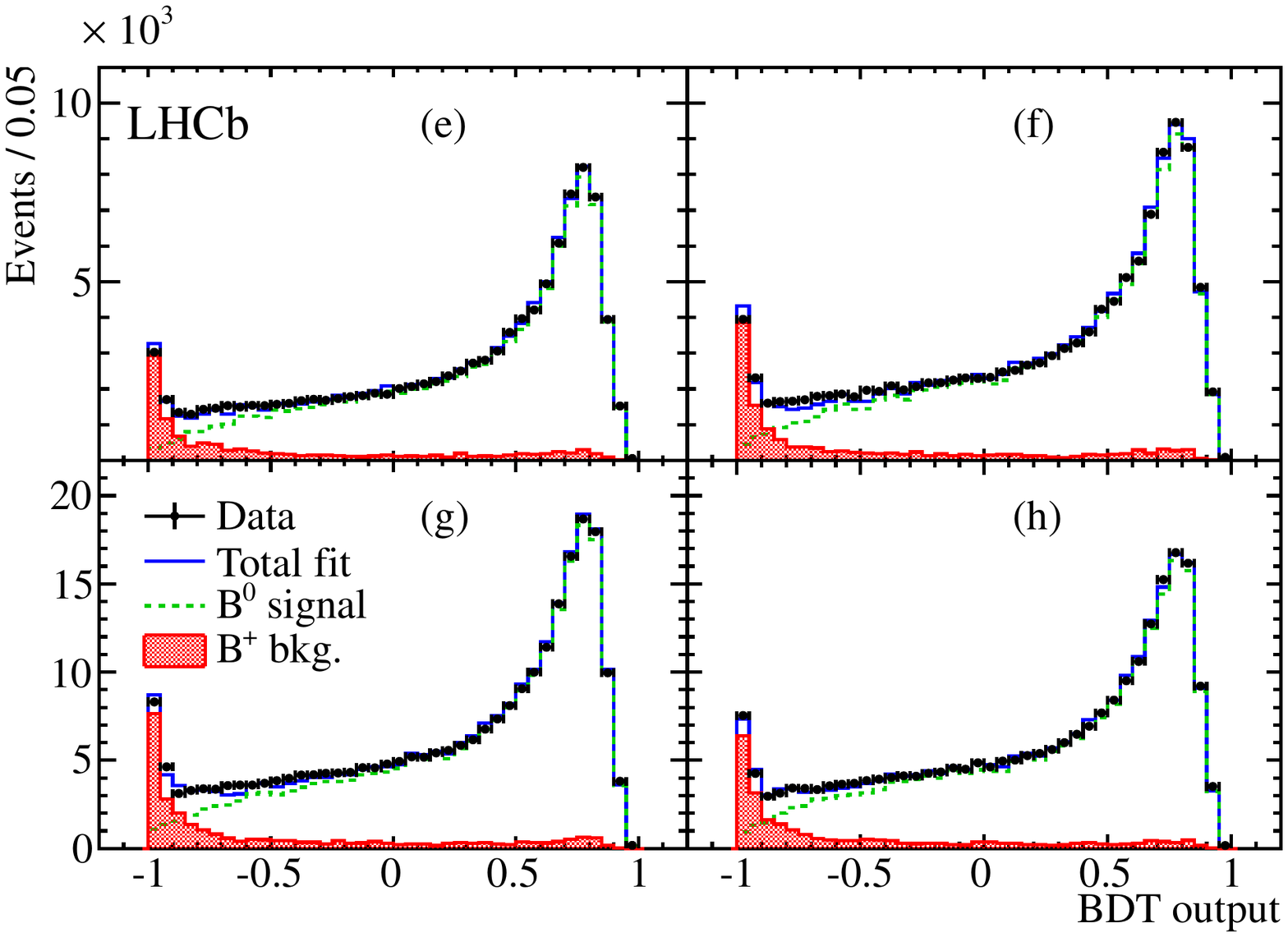}
    \caption{\small{Fits to the output of the \Bu veto BDT for
        (top four plots) \DpMode and (bottom four plots) \DstarMode in 2012 data, for each tagging
        category. The filled red
        histogram, the dashed green line, and the continuous blue
        line correspond to background, signal, and total templates,
        respectively. The average mistag fraction per category 
        increases when going from (a) to (d), and (e) to (h).}}
    \label{fig:IsoBDT_Fit_2012} 
  \end{center} 
\end{figure}

The fraction of \Bu background in data, $\alpha_{\Bu}$, 
is determined with good 
precision by fitting the distribution of the BDT classifier, where 
templates for signal and \Bu background are obtained from simulation. 
Fits are performed separately in tagging categories for 2011 and
2012 data, giving fractions of \Bu of 6\% and 3\% on average
for the \DmMode and the \DstarMode modes with relative variation
of the order of 10\% between samples. The results of the fits to 2012 data 
for both modes are given in Fig.~\ref{fig:IsoBDT_Fit_2012}. 
Limited knowledge of the exclusive decays used to build the simulation
templates leads to systematic uncertainties of 0.5\% and 0.4\% on the \Bu
fractions for \DmMode and \DstarMode. 
In the decay time fit, the \Bu fractions are kept fixed. 
The statistical and systematic uncertainties on $\alpha_{\Bu}$
lead to a systematic uncertainty on \dmd, which is reported in Sec.~\ref{sec:systematics}. 

The oscillation frequency \dmd is determined from a binned maximum likelihood fit 
to the distribution of the \Bz
decay time $t$ of candidates classified as mixed ($q=-1$) or unmixed ($q=1$) according to the flavour of the \Bd meson
at production and decay time.

The total PDF for the fit is given by 
\begin{equation}\label{eq:pdf_time_sFIT}
 {\cal P}(t,q)      =  { \cal S}(t,q)  + \alpha_{_\Bu} {\cal B}^{+}(t,q)  \, ,
\end{equation}
\noindent where the time distributions for signal and background are given by 
\begin{eqnarray}
  \label{eq:pdf_time}
   {\cal S}(t,q)  & = & {\cal N} e^{-\Gd t}\Big( 1 + q (1 -
                        2\mistag_{\rm sig}) \cos\dmd t \Big) \, ,\\
  {\cal B^+}(t,q) & = & {\cal N}_{_\Bu} e^{-\Gamma_u
                        t}\left(\frac{1+q}{2}-q \mistag_{\Bu}\right)
                        \, . \nonumber
\end{eqnarray}
\noindent Here ${\cal N}$ and ${\cal N_{_\Bu}}$ are 
normalisation factors, and $\Gamma_d$ and $\Gamma_u$  are fixed in the
fit to their world average values~\cite{PDG2014}, where $\Gamma_u = 1/{\tau_{\Bu}}$, 
with $\tau_{\Bu}$ being the lifetime of the \Bu meson. The mistag fractions for signal and \Bu components,
$\mistag_{\rm sig}$ and $\mistag_{\Bu}$, vary freely in the fit.
To account for the time resolution, both distributions in
Eq.~\ref{eq:pdf_time} are convolved with a 
resolution model
that takes into account uncertainties on both the decay length and the momentum.
The distributions used in the fit are therefore obtained by a double convolution. 
The contribution accounting for the decay length resolution is described by a triple Gaussian function with an effective width
corresponding to a time resolution of 
$75\fs$, as determined from simulation. The contribution accounting for the uncertainty on the momentum is described by 
the distribution of $p_{\rm rec}/(k \cdot
p_{\rm true})$, obtained from the simulation. 
This second convolution is  dominant above 1.5 \ps.
Finally, the function $\cal{P}$ is multiplied by an acceptance function
$a(t)$ to account for the effect of the trigger and offline selection and
reconstruction. 
The acceptance is described by a sum of { cubic spline}
polynomials~\cite{splines}, which may be different for signal and \Bu
background. The ratios between spline coefficients of the \Bu 
background acceptance and those of the signal acceptance are fixed to
the values predicted by simulation. The spline coefficients for signal
are then determined for each tagging category directly from the tagged 
time-dependent fit to data. 

The fitting strategy is validated with simulation. A bias is observed 
in the \dmd value, due to a correlation between the
decay time and its resolution,  which is not taken into account when parameterizing the
signal shape. Simulation shows that this correlation is introduced by the requirements of the software
trigger and offline selection 
on the impact parameters of \Dm and \Dzb with
respect to the PV. 
Values for this bias, of up to 4\invns with a 10\% uncertainty, are
determined for each mode and for each year by 
fitting the true and corrected time distributions 
and taking the differences between the resulting values of \dmd. 
The uncertainty on the
bias is treated as a systematic uncertainty on \dmd.

\begin{figure}[b!]
  \begin{center} 
    \includegraphics[width=0.49\textwidth]{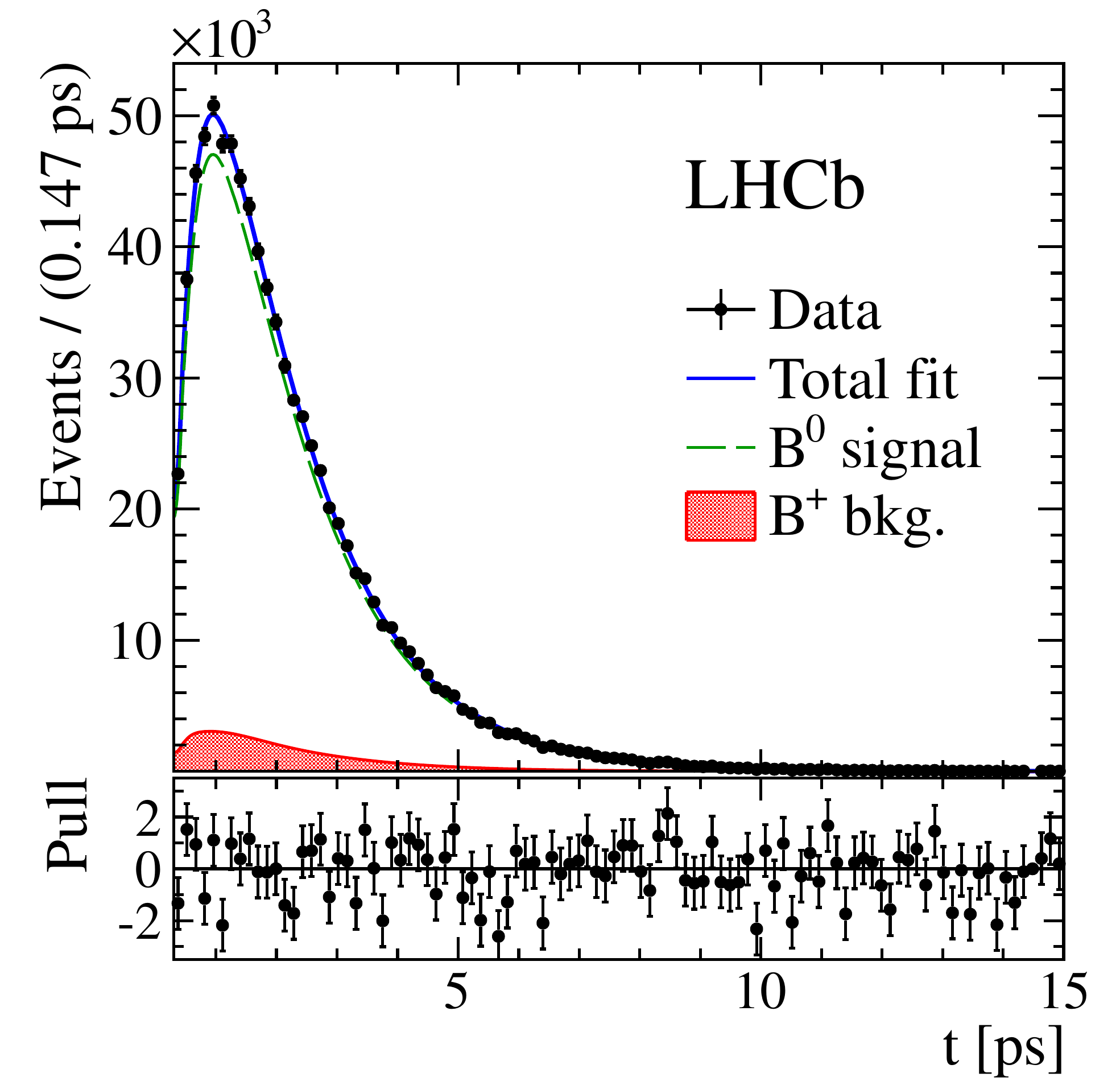}
    \includegraphics[width=0.49\textwidth]{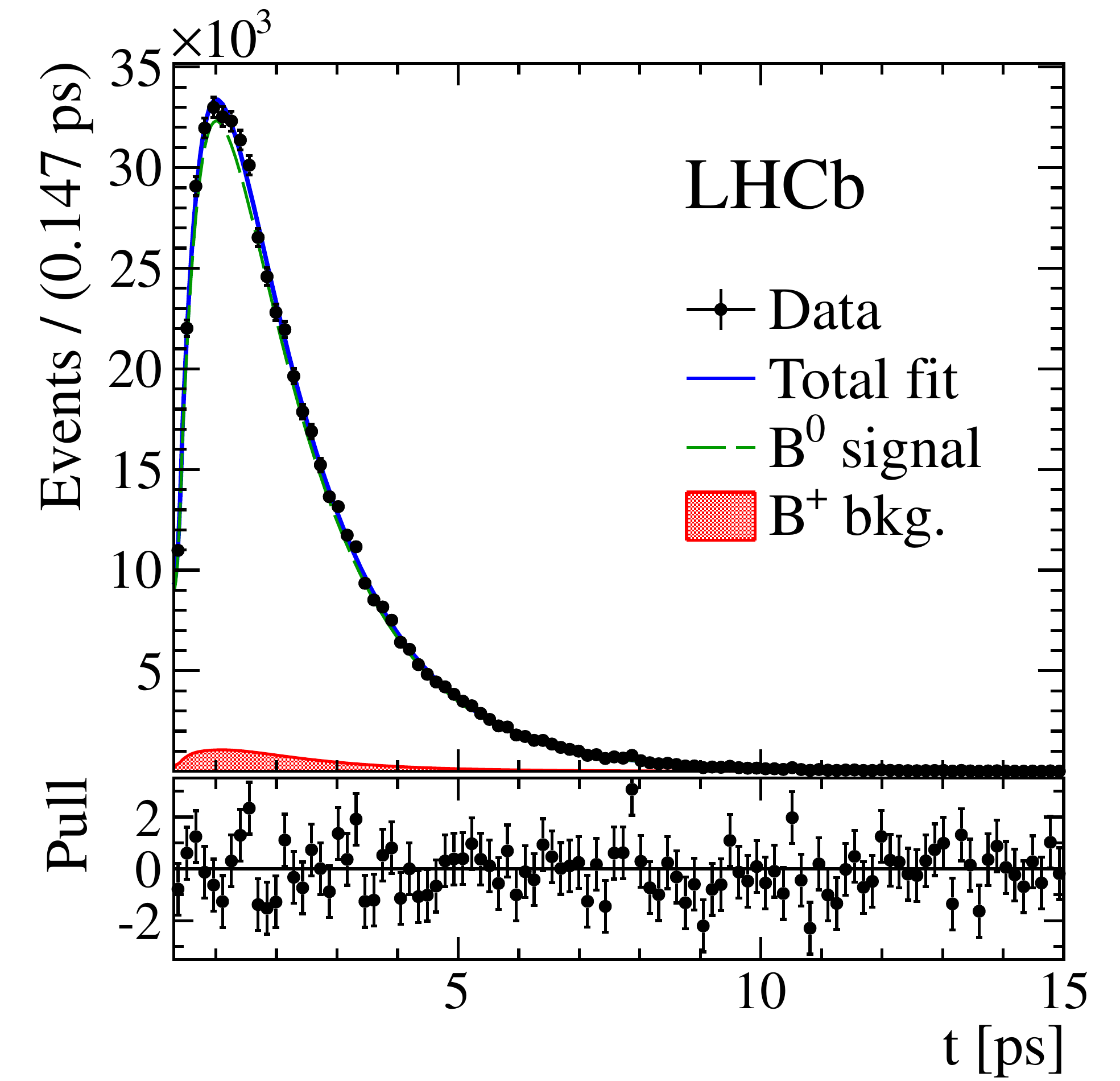}
    \caption{\small{Decay time distributions for (left) \DmMode and
        (right) \DstarMode in the category with lowest mistag in 2012
        data.}}
    \label{fig:time_example} 
  \end{center} 
\end{figure}
\begin{figure}
  \begin{center} 
   \includegraphics[width=0.92\textwidth]{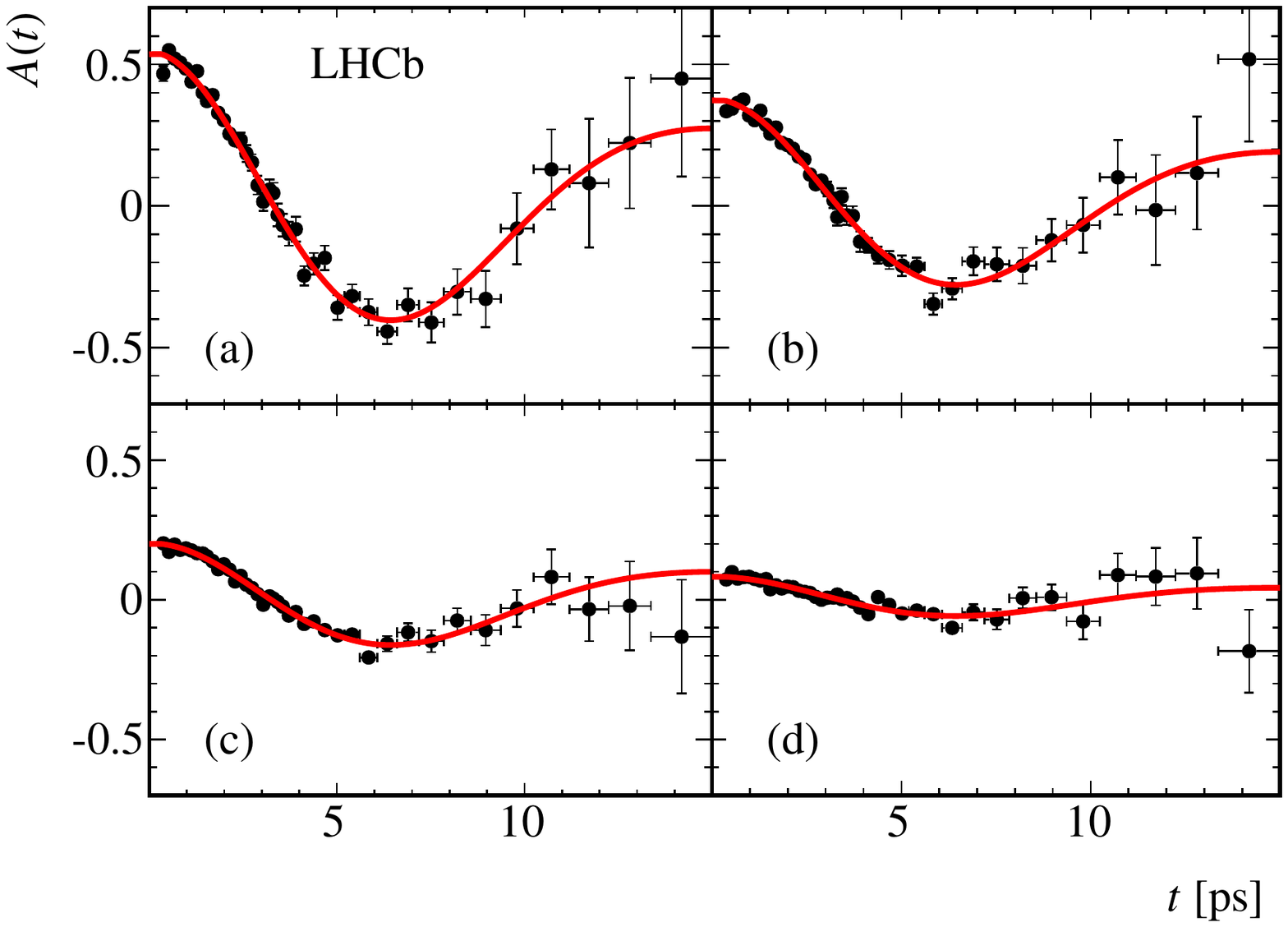}
   \includegraphics[width=0.92\textwidth]{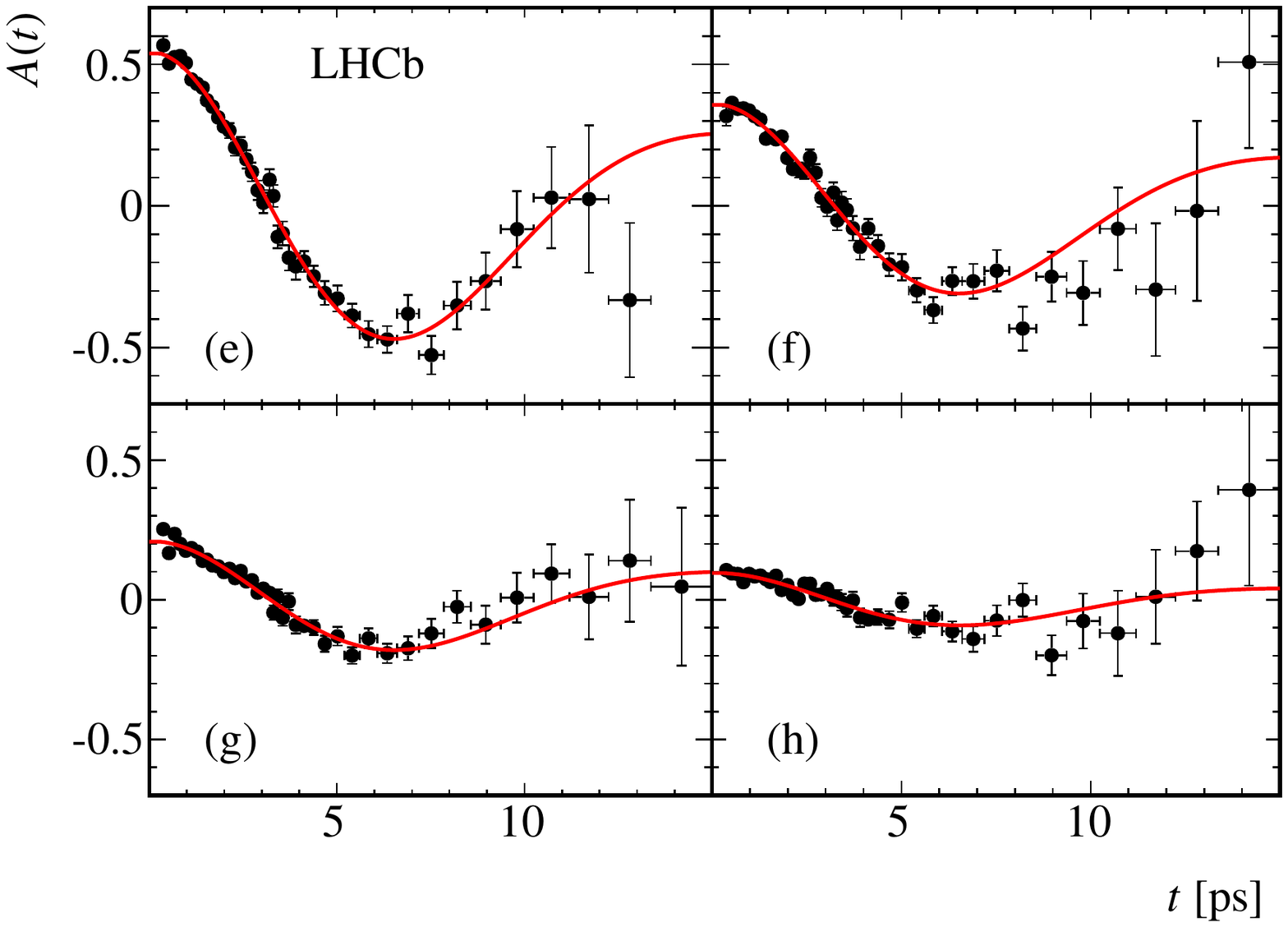}
    \caption{\small{Mixing asymmetry projections in the four tagging
     categories for (top plots) \DmMode  and (bottom plots) \DstarMode for 2011 data.
     The average mistag per category increases when going from (a)
     to (d), and from (e) to (h).}}
     \label{fig:proj_asym_cat_2011} 
  \end{center} 
\end{figure}
\begin{figure}
  \begin{center} 
   \includegraphics[width=0.92\textwidth]{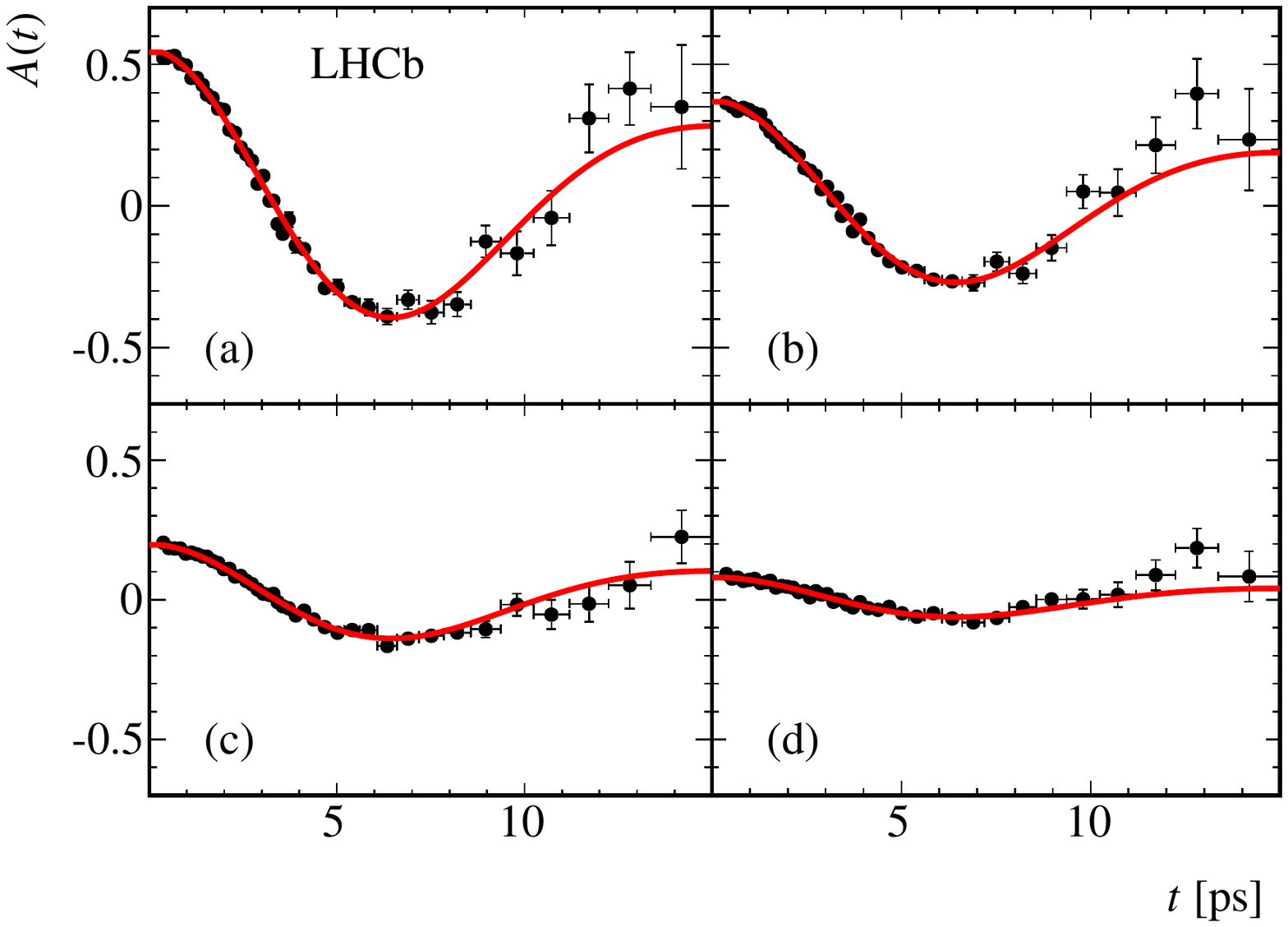}
   \includegraphics[width=0.92\textwidth]{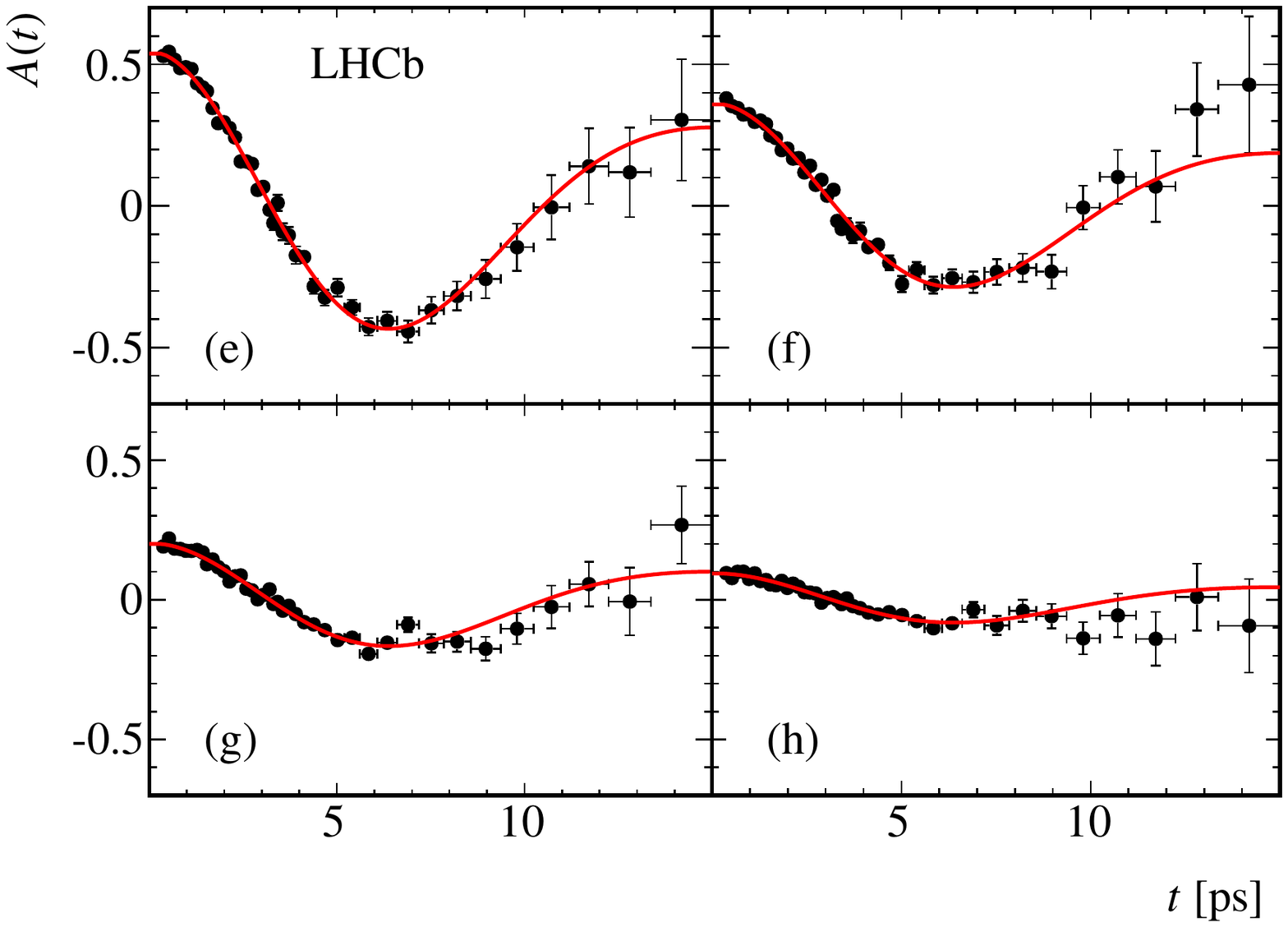}
    \caption{\small{Mixing asymmetry projections in the four tagging
     categories for (top plots) \DmMode  and (bottom plots) \DstarMode for 2012 data.
     The average mistag per category increases when going from (a)
     to (d), and from (e) to (h).}}
     \label{fig:proj_asym_cat_2012} 
  \end{center} 
\end{figure}

The values of \dmd, obtained from  the time-dependent fit and corrected
for the fit bias, are reported in
Table~\ref{tab:FitResults_dmd}. Systematic uncertainties are
discussed below. 
The four independent \dmd values are compatible within
statistical uncertainties. 
Figure~\ref{fig:time_example} shows the fit
projections for the decay time distributions for the candidates in the
category with lowest mistag rate in 2012 data. 
The time-dependent asymmetries for the \DmMode and \DstarMode modes in 2011 and 2012
data are shown in Figs.~\ref{fig:proj_asym_cat_2011} and~\ref{fig:proj_asym_cat_2012}. 
Fits are also performed in subsamples of different track multiplicity, number of primary vertices, magnet polarity, run periods, and muon charges. 
Statistically compatible results are obtained in all cases. 
A combination of the two \dmd determinations, including systematic uncertainties, is given in Sec.~\ref{sec:conclusion}. 

\begin{table}[bt]
  \begin{center}
    \caption{\small Results for \dmd measured in each mode for 2011 and 2012 data separately,
      for the total sample, and for the combination of the two modes. 
The quoted uncertainties for the separate samples are statistical
only. For the total samples and the combination, they refer to statistical and total systematic uncertainties, respectively.}
    \label{tab:FitResults_dmd}
    \begin{tabular}{lccc}
      Mode & 2011 sample & 2012 sample & Total sample \\
               &         \dmd [\invns]& \dmd [\invns] & \dmd [\invns] \\
      \hline
      \DpMode                    &  $506.2\pm5.1$  & $505.2\pm3.1$ & $505.5\pm2.7\pm1.1$\\
      \DstarMode                 &  $497.5\pm6.1$  & $508.3\pm4.0$ & $504.4\pm3.4\pm1.0$\\
      \hline
            combination      & & & {$505.0\pm2.1\pm1.0$} \\

    \end{tabular}
  \end{center}
\end{table}

\section{Systematic uncertainties}
\label{sec:systematics}

The contribution of each source of systematic uncertainty is evaluated 
by using a large number of parameterized simulations.
The difference between the default \dmd value and the result obtained when repeating the fits after
having adjusted the inputs to those corresponding to the systematic
variation under test, is taken as a systematic uncertainty.
Systematic uncertainties are summarized in
Table~\ref{tab:sys_overall_summary}. 
\begin{table}[t]
\caption {\small Sources of systematic uncertainties on \dmd, separated into 
  those that are correlated and uncorrelated between the two decay channels \DmMode and \DstarMode. }
  \label{tab:sys_overall_summary}
  \vspace{0.1cm}
  \begin{center}
      {\small
        \begin{tabular}{l cc cc}
          \multirow{2}{*}{Source of uncertainty}  & \multicolumn{2}{c}{\DmMode [\invns]}  & \multicolumn{2}{c}{\DstarMode [\invns]}  \\
                                             & Uncorrelated & Correlated  & Uncorrelated &  Correlated  \\
          \hline
             \Bu\ background                 & 0.4  & 0.1  &  0.4  & --  \\
             Other backgrounds               & --   & 0.5  & --    & --  \\
             $k$-factor distribution         & 0.4  & 0.5  &  0.3  & 0.6  \\
             Other fit-related               & 0.5  & 0.4  &  0.3  & 0.5  \\ 
              \hline
              Total                          & 0.8  & 0.8 &  0.6 & 0.8 \\
        \end{tabular}
      }
  \end{center}
\end{table}

\subsection{Background from $\mathbf{B^+}$}
The fraction of \Bu background is estimated from data with a very
small statistical uncertainty. 
A variation, within their uncertainties, of the branching fractions of semileptonic \Bd decays
resulting in a \Dstarm or \Dm in the final state gives systematic uncertainties 
on the \Bu fractions of 0.5\% and 0.4\% for \DmMode and \DstarMode. 
The resulting uncertainty on \dmd
is 0.1\invns in \DpMode and is negligible for
\DstarMode. 
In the default fit, the decay time acceptance 
ratio of the \Bz and the \Bu components is taken from
simulation. The time acceptance is to a large extent due to the
cut on the \Dz impact parameter. A possible systematic effect due to an
incorrect determination of the acceptance ratio from simulation is 
estimated by fitting events, generated with the default signal and background
acceptances, with an acceptance ratio determined by using a tighter \Dz IP cut than the default.
This gives an uncertainty of 0.4\invns on both decay
modes. 
The above systematic uncertainties are considered as uncorrelated between the two channels. 

The uncertainty on \dmd from the resolution on the \Bp decay length is
0.1\invns in the \DmMode channel and is negligible in the \DstarMode channel. 

\subsection{Other backgrounds}
The impact of the knowledge of backgrounds due to semileptonic \Bs
decays with \DstDm in the final state is estimated by varying 
their contributions within the uncertainties on their branching fractions. 
This effect has a negligible impact on \dmd for both channels. 
For the \DmMode channel, there is an additional contribution from 
$\Bs\to\Dsm\mup\neum$ decays, where a kaon in the $\Dsm\to\Km\Kp\pim$ decay is
misidentified as a pion, 
which gives an 8\% contribution due to \Dsm peaking under the \Dm mass.  
A difference in \dmd of 0.5\invns is observed. 

The $\Lb\to \neutron\Dstarm\mup \nu_{\mu}$ decay has not been
observed. However, because of the similar final state, 
it can be mistaken for $\Bu$ background, since neither of them exhibits oscillatory behaviour.
Dedicated simulated samples are generated by assuming colour
suppression with respect to signal, and are used to estimate a signal
contamination of 0.2\% from \Lb decays, with $100\%$ uncertainty,
which gives a negligible effect on \dmd. 

Small contributions from $\B\to D^{(*)-} D_s^{+}X$  decays, with the
$\Ds$ decaying semileptonically give an uncertainty of 0.2\invns on
\dmd\ in the \DpMode mode, and a negligible effect for the \DstarMode
mode. 

\subsection{The $\boldmath k$-factor}
Two main sources of systematic uncertainty are related to the $k$-factor. 
The first, due to possible differences in the \PB momentum spectrum
between simulation and data, is studied by comparing the \PB momentum in \BuToJPsiK decays in data
and simulation, and reweighting signal simulation to estimate the effect on the $k$-factor distribution and therefore on \dmd. 
The systematic uncertainties on \dmd from this effect for \DmMode and \DstarMode are 0.3\invns and 0.5\invns. 
The second source, related to the uncertainties on the measurements of the
branching fractions for the exclusive modes which are used to build the
simulated samples, is evaluated by varying the branching fractions of
exclusive decays one at a time by one standard deviation, and reweighting the corresponding $k$-factor distribution.   
An uncertainty of 0.4\invns is obtained for both \DmMode and \DstarMode channels. 
The systematic uncertainties from the $k$-factor
correction are taken to be correlated between the two channels. 

The systematic uncertainties on \dmd from the finite number of events
in the simulation sample used to compute the $k$-factor corrections
are 0.3 and 0.4\invns (\DmMode) and 0.2 and 0.3\invns (\DstarMode) for the 2011 and 2012 samples, respectively.

\subsection{Other systematic uncertainties}
Possible differences between data and simulation in the resolution on the \Bd
flight distance are evaluated by 
using the results of a study reported in Ref.~\cite{LHCb-PAPER-2013-006},
and scaling the widths of the triple Gaussian function by a factor 1.5 with respect to
the default. Uncertainties of 0.3\invns and 0.5\invns on \dmd are obtained for \DmMode and \DstarMode. 
Both channels are affected by the same discrepancy between data and
simulation; thus these systematic uncertainties are taken as correlated. 

Since all parameters are allowed to vary freely in the invariant mass fits, the
uncertainties from the invariant mass model are small. 
As a cross-check, when the fits are repeated using
the {\em{sWeights}} determined without splitting the mass fits in
tagging categories, negligible variation in \dmd is found. 
Signal and background mistag probabilities are free parameters in the fit,
and therefore no systematic uncertainty is associated to them. 

Asymmetries in the production of neutral
and charged \PB mesons, in tagging efficiency and mistag
probabilities, and in the reconstruction of the final state are
neglected in the \dmd fits. Also, the 
\Bd semileptonic \CP asymmetry $a_{\rm sl}^d$ is assumed to be
zero. 
The systematic uncertainty on \dmd arising from these assumptions is studied using parameterized simulations
with the asymmetries set to zero, to their measured values, and to random variations from their central
values within the uncertainties~\cite{LHCb-PAPER-2014-053}. The resulting uncertainty on \dmd is found to be negligible.

The bias in \dmd from the correlation between the decay time and its resolution is determined using the simulation.
The dependence of \dmd on possible differences between data and
simulation has already been 
considered above by varying the composition of the simulation sample used to construct the 
$k$-factor distribution. 
Since the bias is related to the cut on the \D meson IP with
respect to the PV, 
the fits are repeated with a $k$-factor distribution obtained with a tighter cut 
on the IP, and the difference with respect to the
default is taken as the systematic uncertainty. 
The systematic uncertainties (0.5 and 0.3\invns for \DmMode and \DstarMode, respectively) 
related to the bias are considered as uncorrelated between the
channels, as they are 
determined from different simulation samples and the time-biasing cuts, responsible for
the systematic uncertainty on the bias, are different for the two
channels. 

The knowledge of the length scale of the \lhcb experiment is limited by the
uncertainties from the metrology measurements of the
silicon-strip vertex detector. This was evaluated in the
context of the $\Delta m_s$ measurement and found to be
0.022\%~\cite{LHCb-PAPER-2013-006}. This translates into an uncertainty on \dmd 
of 0.1\invns. The uncertainty on the knowledge of the momentum scale is determined
by reconstructing the masses of various particles and is found to be 0.03\%~\cite{LHCb-PAPER-2013-011}.
This uncertainty results in a 0.2 \invns uncertainty in \dmd in both modes. 
Both uncertainties are considered correlated across the two channels.  

Effects due to the choice of the binning scheme and fitting ranges are found to be negligible.

\section{Summary and conclusion}
\label{sec:conclusion}
A combined value of \dmd is obtained as a weighted average of the four
measurements performed in \DmMode and \DstarMode in the years 2011 and 2012. 
First, the 2011 and 2012 results for each decay mode are averaged according to their statistical uncertainties. 
The combined results are shown in the last column of
Table~\ref{tab:FitResults_dmd}. 
Then, the resulting \dmd values of each mode are averaged taking
account of statistical and uncorrelated systematic uncertainties.
The correlated systematic uncertainty is added in quadrature to the resulting uncertainty.
The combined result is shown in the last row of
Table~\ref{tab:FitResults_dmd}.

In conclusion, the oscillation frequency, \dmd, in the \Bd--\Bdb system is
measured in semileptonic \Bd decays using data collected in 2011 and
2012 at \lhcb. 
The decays \DmMode and \DstarMode are used, where the \D mesons are
reconstructed in Cabibbo-favoured decays \DmToKpipi and \DstarmToDpi,
with \DzbToKpi.
A combined \dmd measurement is obtained,  
\begin{equation}
  \dmd = \left(505.0 \pm 2.1 \mathrm{\,\left(stat\right)}\xspace \pm 1.0 \mathrm{\,\left(syst\right)}\xspace\right) \invns \ , \nonumber
\end{equation}
\noindent which is compatible with previous \lhcb results and the world average~\cite{PDG2014}. 
This is the most precise single measurement of this quantity, with a
total uncertainty similar to the current world average.

\section*{Acknowledgements}
\noindent We express our gratitude to our colleagues in the CERN
accelerator departments for the excellent performance of the LHC. We
thank the technical and administrative staff at the LHCb
institutes. We acknowledge support from CERN and from the national
agencies: CAPES, CNPq, FAPERJ and FINEP (Brazil); NSFC (China);
CNRS/IN2P3 (France); BMBF, DFG and MPG (Germany); INFN (Italy); 
FOM and NWO (The Netherlands); MNiSW and NCN (Poland); MEN/IFA (Romania); 
MinES and FANO (Russia); MinECo (Spain); SNSF and SER (Switzerland); 
NASU (Ukraine); STFC (United Kingdom); NSF (USA).
We acknowledge the computing resources that are provided by CERN, IN2P3 (France), KIT and DESY (Germany), INFN (Italy), SURF (The Netherlands), PIC (Spain), GridPP (United Kingdom), RRCKI and Yandex LLC (Russia), CSCS (Switzerland), IFIN-HH (Romania), CBPF (Brazil), PL-GRID (Poland) and OSC (USA). We are indebted to the communities behind the multiple open 
source software packages on which we depend.
Individual groups or members have received support from AvH Foundation (Germany),
EPLANET, Marie Sk\l{}odowska-Curie Actions and ERC (European Union), 
Conseil G\'{e}n\'{e}ral de Haute-Savoie, Labex ENIGMASS and OCEVU, 
R\'{e}gion Auvergne (France), RFBR and Yandex LLC (Russia), GVA, XuntaGal and GENCAT (Spain), Herchel Smith Fund, The Royal Society, Royal Commission for the Exhibition of 1851 and the Leverhulme Trust (United Kingdom).

\clearpage

\appendix
\section{Appendix}
\label{sec:Supplementary-App}

\subsection{BDT classifier}
 The variables used as input for the BDT classifier are the following:
\begin{itemize}
\item Visible mass of the \PB candidate, $m_{\PB} \equiv m(D^{(*)-}\mup)$
\item Corrected mass~\cite{Abe:1997sb}, defined as
  $m_{\rm corr}=\sqrt{m_{\PB}^{2} + p_{T}(\PB)^{2}} +p_{T}(\PB)$, where
 $p_{T}(\PB)$ is the visible momentum of the \PB candidate transverse
 to its flight direction; the \PB flight
  direction is measured using the primary vertex and \PB vertex
  positions  
\item Angle 
between the visible momentum of the \PB  candidate and its flight direction 
\item Impact parameter, ${\rm{IP}}(\pi, D)$, with respect to the decay 
  vertex of the \Dm (\Dzb), of the track with the  
smallest impact parameter with respect to the  \PB candidate 
\item Smallest vertex $\chisq$ 
of the combination of the \Dm (\Dstarm) with any other track, and the invariant mass of
this combination 
\item Cone isolation $I = \frac{p_T(\PB)}{p_T(\PB) + \sum_{i}p_{T,i}}$,
  where the sum is computed over tracks which 
satisfy $\sqrt{\delta \eta_i^{2} +\delta \phi_i^{2}}<1 $, $\delta \eta_i$
and $\delta \phi_i$ being the difference in pseudorapidity and in
polar angle $\phi$ between the track and the \PB candidate 
\item Track isolation variables, used to discriminate tracks
  originating from the \B vertex from those originating 
  elsewhere: 
  \begin{itemize}
  \item Number of nearby tracks~\cite{LHCb-PAPER-2014-052}, computed
    for each track in the \PB decay chain 
  \item The output of an isolation BDT~\cite{LHCb-PAPER-2014-052} estimated for the
    \PB candidate 
\item A second isolation BDT, similar to the previous, which
  exploits a different training strategy and additional variables, 
computed for tracks originating from \Dm (\Dzb) decays, 
those coming from the \PB decay, 
and all tracks in the decay chain. 
  \end{itemize}
\end{itemize}
The TMVA package~\cite{Speckmayer:2010zz}, used to train and test
the classifier, ranks the input variables according to their
discriminating power between signal and
background.

\subsection{Distributions of the $k$-factor}
\begin{figure}
\begin{center} 
  \includegraphics[width=0.59\textwidth]{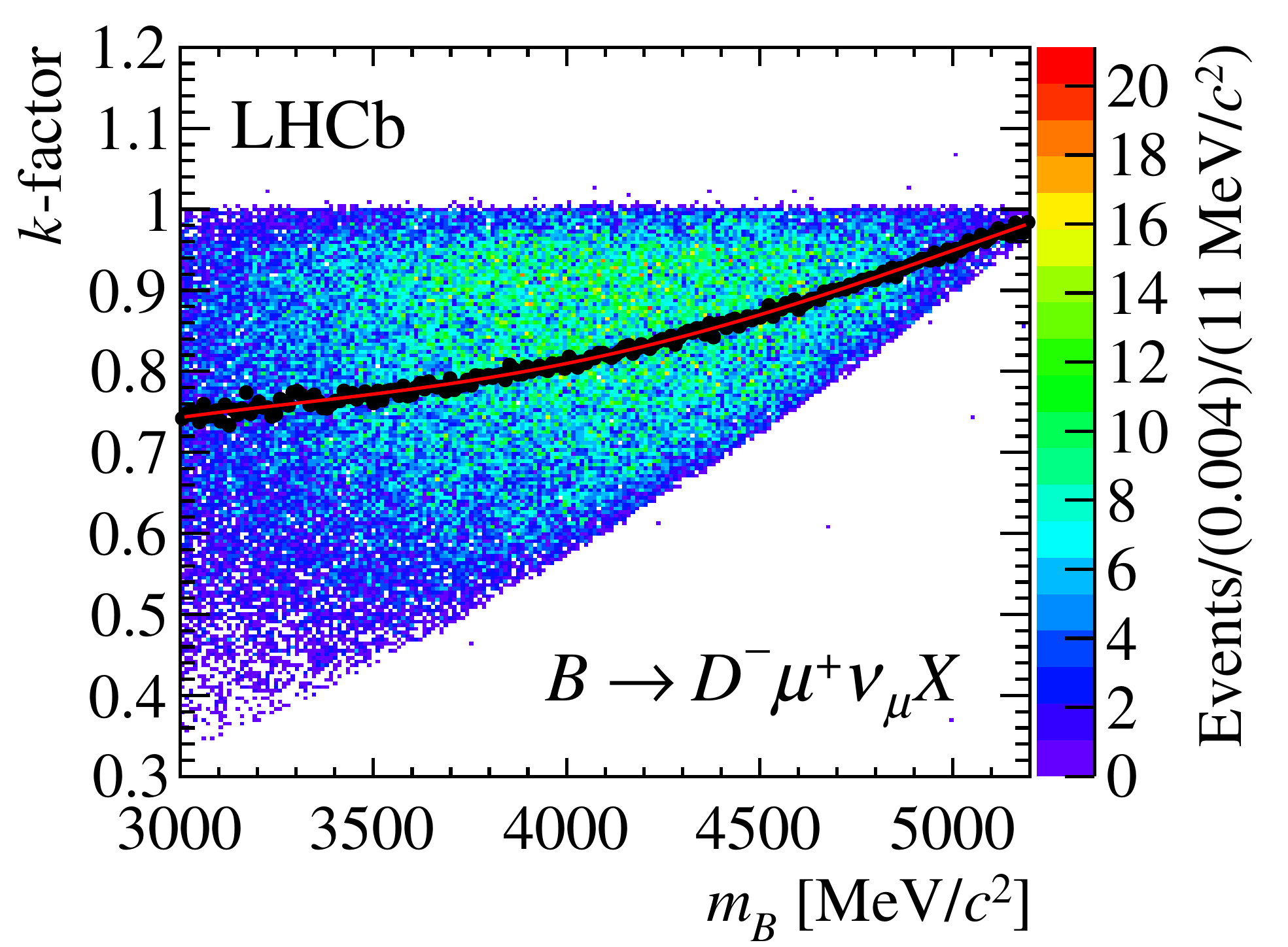}\\
  \includegraphics[width=0.59\textwidth]{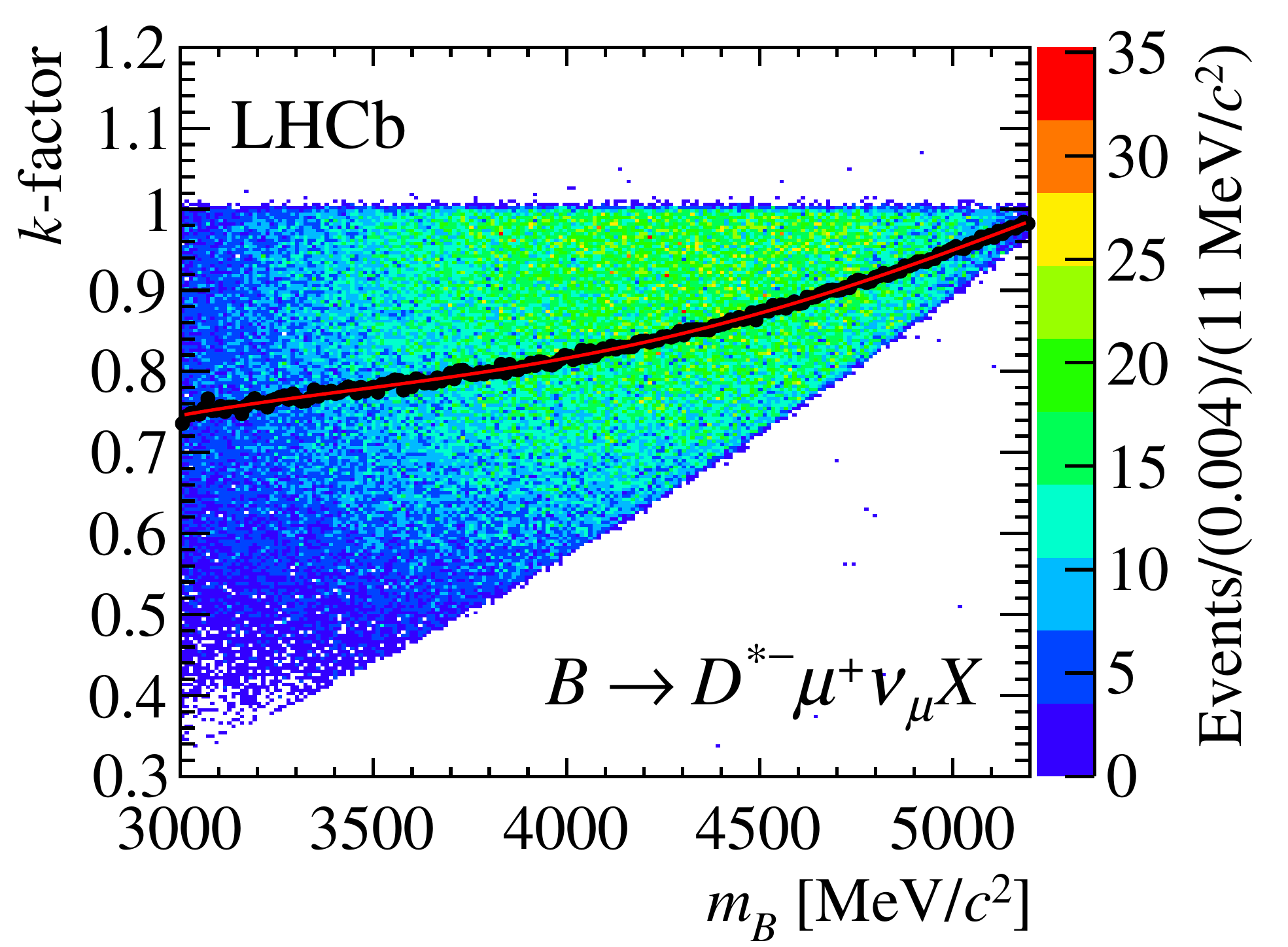}
 \caption{\small The $k$-factor distribution and the average
   $k$-factor (black points) as a function of the
   visible mass of the \PB candidate, in samples of simulated (top) \DpMode and (bottom)
   \DstarMode decays. 
   Polynomial fits to the average $k$-factor  are also shown
   as a solid (red) line.}
\label{fig:kFactorVsMass} 
\end{center} 
\end{figure}

Figure~\ref{fig:kFactorVsMass} shows distributions of the $k$-factor as a
function of the visible mass of the \PB candidate, as
obtained with samples of simulated signal events. 
In each plot, the average $k$-factor and the result of a
polynomial fit are also shown. 

\clearpage

\addcontentsline{toc}{section}{References}
\setboolean{inbibliography}{true}

\bibliographystyle{LHCb}
\bibliography{main}

\newpage

\centerline{\large\bf LHCb collaboration}
\begin{flushleft}
\small
R.~Aaij$^{39}$, 
C.~Abell\'{a}n~Beteta$^{41}$, 
B.~Adeva$^{38}$, 
M.~Adinolfi$^{47}$, 
A.~Affolder$^{53}$, 
Z.~Ajaltouni$^{5}$, 
S.~Akar$^{6}$, 
J.~Albrecht$^{10}$, 
F.~Alessio$^{39}$, 
M.~Alexander$^{52}$
% ...

{\footnotesize \it
% ...
$ ^{5}$Clermont Universit\'{e}, Universit\'{e} Blaise Pascal, CNRS/IN2P3, LPC, Clermont-Ferrand, France\\
$ ^{6}$CPPM, Aix-Marseille Universit\'{e}, CNRS/IN2P3, Marseille, France\\
% ...
$ ^{10}$Fakult\"{a}t Physik, Technische Universit\"{a}t Dortmund, Dortmund, Germany\\
% ...
$ ^{38}$Universidad de Santiago de Compostela, Santiago de Compostela, Spain\\
$ ^{39}$European Organization for Nuclear Research (CERN), Geneva, Switzerland\\
% ...
$ ^{41}$Physik-Institut, Universit\"{a}t Z\"{u}rich, Z\"{u}rich, Switzerland\\
% ...
$ ^{47}$H.H. Wills Physics Laboratory, University of Bristol, Bristol, United Kingdom\\
% ...
$ ^{52}$School of Physics and Astronomy, University of Glasgow, Glasgow, United Kingdom\\
$ ^{53}$Oliver Lodge Laboratory, University of Liverpool, Liverpool, United Kingdom\\
% ...
}
\end{flushleft}

\end{document}